\documentclass[twocolumn]{bmcart}% uncomment this for twocolumn layout and comment line below
\usepackage{amsmath,amsfonts,amssymb}
\usepackage{cite}
\usepackage{ragged2e} 
\RequirePackage{hyperref}
\usepackage[utf8]{inputenc} %unicode support
\usepackage{graphicx}
\graphicspath{{./img/}}

\startlocaldefs
\endlocaldefs

\begin{document}

%%% Start of article front matter
\begin{frontmatter}

\begin{fmbox}
\dochead{Research}


\title{Studying Bioluminescence Flashes with the ANTARES Deep Sea Neutrino Telescope}

%%%%%%%%%%%%%%%%%%%%%%%%%%%%%%%%%%%%%%%%%%%%%%
%%                                          %%
%% Enter the authors here                   %%
%%                                          %%
%% Specify information, if available,       %%
%% in the form:                             %%
%%   <key>={<id1>,<id2>}                    %%
%%   <key>=                                 %%
%% Comment or delete the keys which are     %%
%% not used. Repeat \author command as much %%
%% as required.                             %%
%%                                          %%
%%%%%%%%%%%%%%%%%%%%%%%%%%%%%%%%%%%%%%%%%%%%%%

\author[
  addressref={IFT, TUM},                   % id's of addresses, e.g. {aff1,aff2}
  corref={TUM},
  email={nico.reeb@tum.de}   % email address
]{\inits{N.R.}\fnm{N.} \snm{Reeb}}
\author[
  addressref={IFT, IMAPP},
  noteref={n1},
  email={hutsch@astro.ru.nl}
]{\inits{S.H.}\fnm{S.} \snm{Hutschenreuter}}
\author[
  addressref={IFT, LMU, CERN},
  noteref={n1},
  email={philipp.zehetner@cern.ch}
]{\inits{P.Z.}\fnm{P.} \snm{Zehetner}}
\author[
  addressref={IFT, LMU},
  noteref={n1},
  email={ensslin@mpa-garching.mpg.de}
]{\inits{T.E.}\fnm{T.} \snm{Ensslin}}
%\author[addressref={Erlangen}]{T.~Eberl}
%\author{and the ANTARES~Collaboration}
\textsf{and}

\author[addressref={IPHC,UHA}]{A.~Albert}
\author[addressref={IFIC}]{S.~Alves}
\author[addressref={UPC}]{M.~Andr\'e}
\author[addressref={Genova}]{M.~Anghinolfi}
\author[addressref={Erlangen}]{G.~Anton}
\author[addressref={UPV}]{M.~Ardid}
\author[addressref={CPPM}]{J.-J.~Aubert}
\author[addressref={APC}]{J.~Aublin}
\author[addressref={APC}]{B.~Baret}
\author[addressref={LAM}]{S.~Basa}
% ...
\textsf{(ANTARES Collaboration)}

%%%%%%%%%%%%%%%%%%%%%%%%%%%%%%%%%%%%%%%%%%%%%%
%%                                          %%
%% Enter the authors' addresses here        %%
%%                                          %%
%% Repeat \address commands as much as      %%
%% required.                                %%
%%                                          %%
%%%%%%%%%%%%%%%%%%%%%%%%%%%%%%%%%%%%%%%%%%%%%%
\begin{artnotes}
    \note[id=n1]{Equal contributor} 
\end{artnotes}

% ...
\address[id=IPHC]{\scriptsize{Universit\'e de Strasbourg, CNRS,  IPHC UMR 7178, F-67000 Strasbourg, France}}
\address[id=UHA]{\scriptsize Universit\'e de Haute Alsace, F-68200 Mulhouse, France}
\address[id=IFIC]{\scriptsize{IFIC - Instituto de F\'isica Corpuscular (CSIC - Universitat de Val\`encia) c/ Catedr\'atico Jos\'e Beltr\'an, 2 E-46980 Paterna, Valencia, Spain}}
\address[id=UPC]{\scriptsize{Technical University of Catalonia, Laboratory of Applied Bioacoustics, Rambla Exposici\'o, 08800 Vilanova i la Geltr\'u, Barcelona, Spain}}
\address[id=Genova]{\scriptsize{INFN - Sezione di Genova, Via Dodecaneso 33, 16146 Genova, Italy}}
\address[id=Erlangen]{\scriptsize{Friedrich-Alexander-Universit\"at Erlangen-N\"urnberg, Erlangen Centre for Astroparticle Physics, Erwin-Rommel-Str. 1, 91058 Erlangen, Germany}}
\address[id=UPV]{\scriptsize{Institut d'Investigaci\'o per a la Gesti\'o Integrada de les Zones Costaneres (IGIC) - Universitat Polit\`ecnica de Val\`encia. C/  Paranimf 1, 46730 Gandia, Spain}}
\address[id=CPPM]{\scriptsize{Aix Marseille Univ, CNRS/IN2P3, CPPM, Marseille, France}}
\address[id=APC]{\scriptsize{Universit\'e de Paris, CNRS, Astroparticule et Cosmologie, F-75013 Paris, France}}
\address[id=LAM]{\scriptsize{Aix Marseille Univ, CNRS, CNES, LAM, Marseille, France }}
% ...
\address[id=Genova-UNI]{\scriptsize{Dipartimento di Fisica dell'Universit\`a, Via Dodecaneso 33, 16146 Genova, Italy}}
% ...

%%%%%%%%%%%%%%%%%%%%%%%%%%%%%%%%%%%%%%%%%%%%%%
%%                                          %%
%% Enter short notes here                   %%
%%                                          %%
%% Short notes will be after addresses      %%
%% on first page.                           %%
%%                                          %%
%%%%%%%%%%%%%%%%%%%%%%%%%%%%%%%%%%%%%%%%%%%%%%

%%\note{Sample of title note}     % note to the article
%%\note[id=n1]{Equal contributor} % note, connected to author

%\end{fmbox}% comment this for two column layout

%%%%%%%%%%%%%%%%%%%%%%%%%%%%%%%%%%%%%%%%%%%%%%%
%%                                           %%
%% The Abstract begins here                  %%
%%                                           %%
%% Please refer to the Instructions for      %%
%% authors on https://www.biomedcentral.com/ %%
%% and include the section headings          %%
%% accordingly for your article type.        %%
%%                                           %%
%%%%%%%%%%%%%%%%%%%%%%%%%%%%%%%%%%%%%%%%%%%%%%%

\vspace{-0.5cm}
\begin{abstractbox}
\begin{abstract} % abstract
We develop a novel technique to exploit the extensive data sets provided by
underwater neutrino telescopes to gain information on bioluminescence in the
deep sea.  The passive nature of the telescopes gives us the unique opportunity
to infer information on bioluminescent organisms without actively interfering
with them.  We propose a statistical method that allows us to reconstruct the
light emission of individual organisms, as well as their location and movement.
A mathematical model is built to describe the measurement process of underwater
neutrino telescopes and the signal generation of the biological organisms. The
Metric Gaussian Variational Inference algorithm is used to reconstruct the model
parameters using photon counts recorded by the neutrino detectors. We apply this
method to synthetic data sets and data collected by the ANTARES neutrino
telescope.  The telescope is located 40 km off the French coast and fixed to the
sea floor at a depth of 2475 m.  The runs with synthetic data reveal that we can
reliably model the emitted bioluminescent flashes of the organisms. Furthermore,
we find that the spatial resolution of the localization of light sources highly
depends on the configuration of the telescope. Precise measurements of the
efficiencies of the detectors and the attenuation length of the water are
crucial to reconstruct the light emission. Finally, the application to ANTARES
data reveals the first precise localizations of bioluminescent organisms using
neutrino telescope data.%
%known to us
\end{abstract}

%%%%%%%%%%%%%%%%%%%%%%%%%%%%%%%%%%%%%%%%%%%%%%
%%                                          %%
%% The keywords begin here                  %%
%%                                          %%
%% Put each keyword in separate \kwd{}.     %%
%%                                          %%
%%%%%%%%%%%%%%%%%%%%%%%%%%%%%%%%%%%%%%%%%%%%%%

%\kwd[Primary ]{}
%\kwd{}
%\kwd[; secondary ]{}
%\end{keyword}

\end{abstractbox}

\end{fmbox}% uncomment this for two column layout

\end{frontmatter}

%%%%%%%%%%%%%%%%%%%%%%%%%%%%%%%%%%%%%%%%%%%%%%%%
%%                                            %%
%% The Main Body begins here                  %%
%%                                            %%
%% Please refer to the instructions for       %%
%% authors on:                                %%
%% https://www.biomedcentral.com/getpublished %%
%% and include the section headings           %%
%% accordingly for your article type.         %%
%%                                            %e
%% See the Results and Discussion section     %%
%% for details on how to create sub-sections  %%
%%                                            %%
%% use \cite{...} to cite references          %%
%%  \cite{koon} and                           %%
%%  \cite{oreg,khar,zvai,xjon,schn,pond}      %%
%%                                            %%
%%%%%%%%%%%%%%%%%%%%%%%%%%%%%%%%%%%%%%%%%%%%%%%%

%%%%%%%%%%%%%%%%%%%%%%%%% start of article main body
% <put your article body there>

%%%%%%%%%%%%%%%%%%%%%%%%%%%%%%%%%%%%%%%%%%%%%%%%%%%%%%%%%%%%%%%%%%%%%%%%%%%%%%%
% INTRODUCTION
%%%%%%%%%%%%%%%%%%%%%%%%%%%%%%%%%%%%%%%%%%%%%%%%%%%%%%%%%%%%%%%%%%%%%%%%%%%%%%%

\section{Introduction}
\label{sec:intro}

The deep sea is one of the remotest habitats on Earth and its biological
diversity is largely unexplored. 76~\% of the inhabitants in the deep pelagic
zone emit light to communicate, attract prey or to protect themselves
\cite{bio:quantification}. The trait of bioluminescence is distributed over a
diverse range of marine species, from bacteria to fish \cite{bio:bioluminocean}.
Over the last years the distribution and quantification of bioluminescence in
the deep sea and individual luminescent organisms have been studied using a
variety of observational techniques \cite{bio:quantification, bio:bathy,
bio:relation,bio:distribution,bio:lightvision}. Most in-situ observation
techniques rely on actively triggering the light production by disturbing the
environment and stimulating the organisms, since spontaneous emission does not
occur at statistically sufficient rates for observation times in the order of
hours\cite{bio:freefall, bio:monterey, bio:meditsea}.  In this context,
spontaneous emission refers to light emission which is not intentionally
stimulated by observers, similar as in \cite{bio:monterey}.  The free-fall
lander observations in the Atlantic Ocean off Cape Verde has detected 5
events per hour when fixing the sensor $250$~m above the sea floor at $4700$~m
leading to no active stimulation due to movements of the sensor
\cite{bio:freefall}.  With these studies, similarities between emitted
bioluminescence flashes have been observed. Most organisms emit a single light
flash or a series of flashes; a rapid increase of the luminosity indicates the
start of a flash which - after reaching its peak value - is decaying
exponentially with a time constant significantly longer than that of the inital
rise. Studies have characterized such flash lightcurves from various species by
the duration, the maximum photon flux and the total photon emission of the flash
\cite{bio:meditsea, bio:bioluminocean, bio:fluor, bio:char1, bio:char2,
bio:char3, bio:char4, bio:char5, bio:char6, bio:char7, bio:char8, bio:char9,
bio:char10, bio:char11, bio:char12, bio:char13, bio:char14, bio:char15,
bio:char16, bio:char18, bio:char19, bio:char20, bio:char21}.  An overview of
these characteristics is provided in table \ref{tab:bio_attr}.  Most species in
the benthic and pelagic zone emit light flashes with their emission maxima
within the range of $\lambda = 450 - 520$~nm \cite{bio:bioluminocean} which
corresponds to the wavelength window of maximum light transmission in seawater
\cite{antares:lighttransmission}.

\begin{table}[!h] 
  \label{tab:bio_attr} 
  \caption{Characteristics of bioluminescent flashes taken from \cite{bio:meditsea, bio:fluor, bio:bioluminocean}.
  } 
  \begin{tabular}{l  l} 
    \hline
    Characteristics & Values \\
    \hline
    Mean duration of flash (s) & $ 0.1 - 59.0$ \\
    Maximum photon flux (photons s$^{-1}$) & $4.9 \cdot 10^{7} - 6.4 \cdot
    10^{12}$ \\
    Total light emission (photons flash$^{-1}$) & $2.4 \cdot 10^8 - 2.3 \cdot
    10^{13}$ \\
    Spectral wavelentgh (nm) & $450 - 520$\\
    \hline
  \end{tabular} 
\end{table}

For years, deep sea neutrino telescopes have been monitoring and recording the
light activity. These telescopes aim to detect Cherenkov radiation caused by
charged secondary particles, which are induced by high-energy cosmic neutrino
interactions with constituents of water molecules. The records of these
telescopes were used to analyze the dynamics of deep sea bioluminescence
\cite{bio:inertial, bio:blooms, bio:relation, bio:vanharen}. The majority of the
recorded bioluminescence is assumed to be triggered by sea currents and
turbulence around the detectors \cite{bio:blooms, bio:relation}.  Therefore, the
challenge to observe spontaneous bioluminescence remains.  In addition to
analysis of long-term temporal changes, the experimental setup and long-term
observations of deep sea neutrino telescopes offers also the possibility to
analyze light emission of individual organisms and relatively rare events of
spontaneous bioluminescence.

In this paper, we present a method to reconstruct the movement and
characteristic lightcurves of individual biological sources in the deep sea with
data of a neutrino telescope. In particular, we use both synthetic, as well as
measured data of the ANTARES neutrino telescope located 40 km off the French
coast on the bottom of the Mediterranean Sea ($42^{\circ}48'$N, $6^{\circ}10'$E)
and anchored to the sea floor at a depth of $2475$~m. The method enabled us to
do the localization of a luminescent organism using ANTARES data and the
simultaneuous reconstruction of the corresponding emitted bioluminescence
lightcurve.

To do so, we developed a generative model of the measurement process of a
neutrino detector and the signal generation of the biological sources to
understand the origin of the measured data. The model parameters such as source
movement and characteristic lightcurves are reconstructed by a Variational
Bayesian Inference algorithm, called Metric Gaussian Variational Inference
(MGVI) \cite{ift:mgvi}. The NIFTy (Numerical Information Field Theory) framework
\cite{ift:niftysoftware, ift:nifty5} provides an implementation of the MGVI
algorithm and has been used to obtain results of this work.

The paper is structured as follows; in section~\ref{sec:data} we summarize
the data provided by the ANTARES experiment and used for the reconstruction.
In section~\ref{sec:methods}, the proposed method is explained by describing
the generative model in detail and highlighting our assumptions. In section~\ref{sec:simulation}, 
simulations and reconstructions on synthetic data sets
are performed to discuss the opportunities and limitations of the model. The
first localizations of deep sea organisms and reconstructions of their
emitted bioluminescence lightcurves using the ANTARES detector are presented
and discussed in section~\ref{sec:analysis}.

\section{Data}
\label{sec:data}

\begin{figure*}[!h] 
  \includegraphics[width=0.97\textwidth]{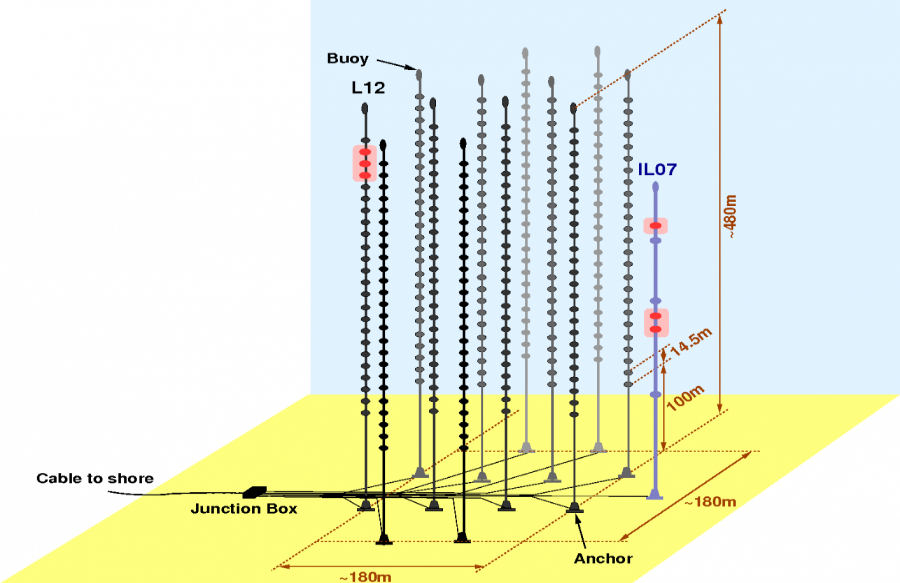}
  \caption{Schematic of the ANTARES setup. \cite{antares:neutrinotelescope}
  }
  \label{fig:antares_setup} 
\end{figure*}

The ANTARES telescope consists of twelve lines that are distributed over an area
of $0.1$~km$^2$. The lines have a length of $480$~m and are placed at a distance
of around $60$ m to each other. Each line, excluding the twelfth, contains 25
storeys with a vertical separation of $14.5$~m between them. The 5 top storeys of
line 12 are not equipped with photomultipliers, but with different acoustic
instruments \cite{antares:positioning}. The first storey of each line is located
around $100$~m above the seabed. A storey is defined as a collection of three
optical modules each containing a photomultiplier tube (PMT). The optical
modules (OMs) are oriented downward looking under $45$~degrees (zenith angle of
$135$~degrees) and with an angle of 120 degrees to each other in horizontal
directions \cite{antares:neutrinotelescope, antares:opticalmodule}. A schematic
view of the ANTARES setup is given in figure~\ref{fig:antares_setup}.

Each optical module measures the light activity in terms of photon hits. The
photon hits recorded in time frames of $\Delta t = 104.858$~ms are directly sent
to shore and the rates of each time frame are calculated \cite{antares:dataacq}.
According to a trigger system the photon hits are stored for a specific period
of time depending on the type of trigger. In the following, only data samples
with detection periods covering fully the bioluminescence lightcurve of interest
are used to analyze the biological behavior. About 1 to 2 of such recordings
are saved per day containing around $2$ minutes of continuous raw data.

In addition to the photon count rates, the position and orientation, as well as
the efficiency of each optical module is monitored precisely \cite{antares:eff,
antares:opticalmodule}. The origin of the internal coordinate system used for
the positioning of the OMs is located in the center of the detector volume. This
internal coordinate system of ANTARES indicates the west-east direction as
x-axis and the south-north direction as y-axis. The vertical direction is given
by the z-axis \cite{antares:opticalmodule, antares:sedimentation}. Within this
work we introduce the naming convention of each optical module as a tuple of
line number and number of optical module $(l, n_\text{om})$. The optical modules
of one line are consecutively numbered starting at the bottom of the line.

\begin{figure*}[!h] 
  \includegraphics{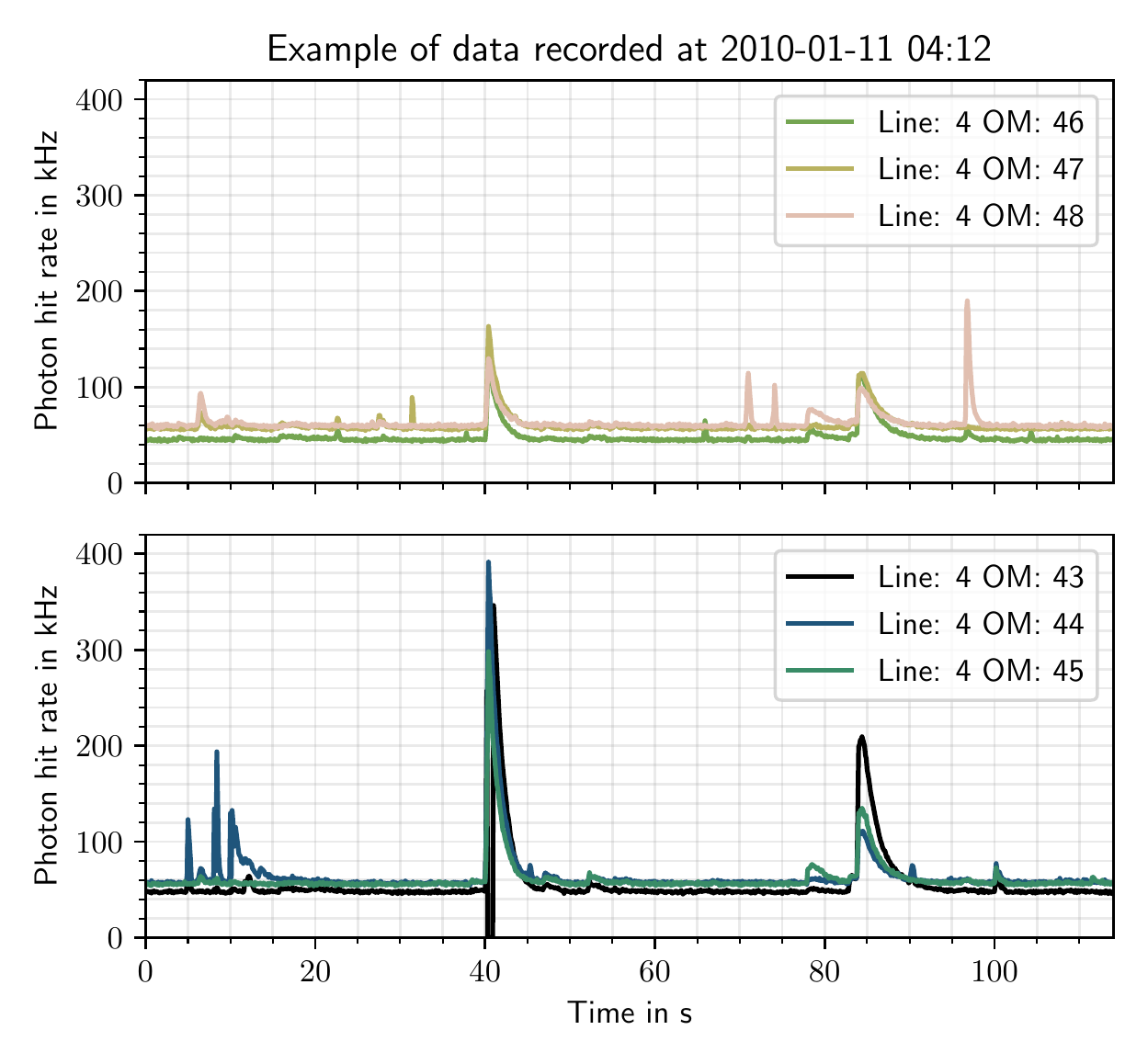}
  \caption{Recorded photon hits of six optical modules (OM). The upper plot
    shows the photon hits of the optical modules in one storey. The lower plot
    shows the photon hits of the optical modules in the storey below.
  } 
  \label{fig:example_data} 
\end{figure*}

An example data file is shown in figure~\ref{fig:example_data}.  In order to
represent the setup of the ANTARES detector, the photon counts of one storey are
grouped together. In the figure, an almost constant background for each optical
module with photon count rates around $40 - 60$~kHz can be identified. This
background is assumed to be induced by $^{40}$K nuclear decays,
bioluminescence, photomultiplier intrinsic noise and radioactive decays in the
sea water and in the glass sphere \cite{antares:backgroundlight}. Furthermore,
we assume to recognize two bioluminescence flashes that are recorded by six
optical modules over two storeys.  The occurrence times of the flashes are $40$~s
and $84$~s after a trigger started the recording. The flash recorded at $t=40$~s
surpassed the threshold limit of optical module $(4, 43)$, i.e. the readout
electronics is saturated which let the recorded photon rate drop to zero for
this detector for a short period. These two flashes are analyzed in section~\ref{sec:analysis}.

The attenuation length of light in the sea water at the ANTARES site was
measured by the ANTARES Collaboration. The light attenuation of water depends on
its chemical and physical properties. Therefore, various measurements of the
water light transmission properties have been made from 1997 to 2010 using an
isotropic light source that emits blue light \cite{antares:lighttransmission,
antares:opticalproperties}. The light attenuation includes the effect of water
absorption of light as well as the impact of scattered photons reaching the
detector \cite{antares:lighttransmission}. A summary of the results of the study
is given in table~\ref{tab:light_att}. Variation in light attenuation depends on
the amount of particles in the water that depends on oceanographic processes, see
\cite{bio:vanharen}. These additional measurements of the state of the detector
and the environment are crucial to perform reasonable reconstructions of
biological light sources.

\begin{table}[!h]
  \caption{Light attenuation length in sea water at ANTARES site
    including only statistical errors \cite{antares:lighttransmission,
    antares:opticalproperties}. 
  } 
  \begin{tabular}{l l}
    \hline
    Measurement & $l_{\text{att}}$ \\
    \hline
    July 1998 & $60.6 \pm 0.4$ m \\
    March 1999 & $51.9 \pm 0.7$ m  \\
    June 2000 & $46.4 \pm 1.9$ m  \\
    May 2008 to March 2010 & $\sim 50 - 60$ m \\
    \hline
  \end{tabular} 
  \label{tab:light_att}
\end{table}

%%%%%%%%%%%%%%%%%%%%%%%%%%%%%%%%%%%%%%%%%%%%%%%%%%%%%%%%%%%%%%%%%%%%%%%%%%%%%%%
% METHODS
%%%%%%%%%%%%%%%%%%%%%%%%%%%%%%%%%%%%%%%%%%%%%%%%%%%%%%%%%%%%%%%%%%%%%%%%%%%%%%%

\section{Methods}
\label{sec:methods}

The data provided by the ANTARES experiment and knowledge about the organisms
and their environment in the deep sea allow us to derive a mathematical model
of the data generation process. This model depends on a set of model parameters $\xi$
and is able to describe a luminescent organism emitting a flash lightcurve,
the propagation of the signal in the deep sea, and the detection of photon
hits at a neutrino detector.

According to Bayes' theorem, measuring photon hits $d$ updates the prior
knowledge on $\xi$, expressed as probability density
function $\mathcal{P}(\xi)$. The resulting posterior density
$\mathcal{P}(\xi|d)$ can be calculated as follows by knowing the likelihood
$\mathcal{P}(d | \xi)$ of the obtained data, given $\xi$, and
the marginal probability of the data $\mathcal{P}(d)$,
\begin{equation}
  \mathcal{P}(\xi | d) = \frac{\mathcal{P}(d | \xi) \,  \mathcal{P}(\xi)}{
  \mathcal{P}(d)}. %= \frac{\mathcal{P}(d , \xi)}{ \mathcal{P}(d)}.
  \label{bayes}
\end{equation}
In case of high dimensionality and complexity of a model the posterior
distribution $\mathcal{P}(\xi|d)$ is often intractable. To overcome this
issue we approximate this distribution with a simpler distribution
$\widetilde{\mathcal{P}}_{\eta}(\xi)$ depending on variational parameter
$\eta$. Within this work the posterior approximation is performed by the
MGVI algorithm \cite{ift:mgvi, ift:encoding}. 

In order to apply Bayes' update rule and perform the posterior approximation
a detailed understanding of the likelihood is crucial. First, we discuss the
measurement process of a photon detector and derive an expression for the
likelihood of multiple optical modules. Second, we build a mathematical
model for the light emission of an organism and the light propagation
through water reaching the photon detector. We explain each aspect of our
model in detail, beginning with the generic formula of the expected photon
number arriving at one optical module.

\subsection{Likelihood and measurement process}
\label{sec:likelihood}
The optical modules of neutrino telescopes detect single photons. Individual
photon hits can be treated as independent events and therefore the photon
detection is a classical Poisson process. Due to the assumption of Poisson
statistics, the photon rate $r_{i, t}$ over the detection window $(t - \Delta t,
t)$ needs to be converted to the total number of photon hits $d_{i,t} = \Delta t
\cdot r_{i, t}$ detected by optical module $i$. 

The likelihood of measuring $d_{i,t}$ photon hits at the
optical module $i$ at time $t$ for a given expected number of photon hits
$\lambda_{i,t}$ can be written as 
\begin{align} 
  \mathcal{P}(d_{i,t} | \lambda_{i,t}) \, 
  &= \, \lambda_{i,t}^{d_{i,t}} \: \frac{e^{-\lambda_{i,t}}}{d_{i,t} !}.  
  \label{likeli_single} 
\end{align}
The expected number of photon hits $\lambda_{i, t}$ is defined as the photon
counts over the fixed detection time $\Delta t$ with. The measurement process
is independent at different times $t, t+\Delta t, ...$ and at different optical modules $i, i+1, ...$. 
Hence, the likelihood of the count data vector $d$ over a time frame $\Delta T = N \cdot \Delta t$ with $N$
discrete time steps and elements $d_{i,t}$ is the direct product of the single measurement likelihoods,
\begin{align} 
  \mathcal{P}(d | \lambda) \, &= \, \prod_i \prod_{t=0}^{N-1} \lambda_{i,t}^{d_{i,t}} \:
\frac{e^{-\lambda_{i,t}}}{d_{i,t} !} = \prod_i \mathcal{P}(d_i, \lambda_i).
  \label{likeli} 
\end{align}

\subsection{Signal generation}
\label{sec:signal}

\subsubsection{Expected photon number}
\label{sec:response}

The number of expected photons depends highly on the light source itself, but
also on specific attributes of the detector and its surroundings.  The
luminescent organisms are modeled as moving point sources with a position
$\vec{x}(t)$ that generate specific time-varying light patterns spreading
isotropically.\footnote{The assumption of isotropy is an approximation, but
for many of the transparent organisms a reasonable one.} The total amount of
photons emitted by a point source over the detection period $(t - \Delta t,
t)$ is described by the function $\mathcal{N}(t)$. Typically $\mathcal{N}$ is
a vector containing the photon numbers of a time frame $\Delta T$ at $N$
discrete times with time steps of
$\Delta t$. The luminosity $\mathcal{L}$ can be calculated as the rate of
emitted photons per second, $\mathcal{L} = \frac{\mathcal{N}}{\Delta t}$.
Therefore, the number of emitted photons $\mathcal{N}$ is given by multiplying
the luminosity with the time step $\Delta t$.

Due to various factors such as detector quality, detector geometry and water
absorption, the amount of photons reaching the optical modules is reduced.
The efficiency $\epsilon_i$ of the optical module $i$ is sampled
($\hookleftarrow$) from a Gaussian model
\begin{align}
  \label{eq:efficiency}
  \epsilon \hookleftarrow \mathcal{G}(\epsilon - \epsilon_\text{ANTARES}, \sigma_\epsilon^2),
\end{align}
with efficiencies $\epsilon_\text{ANTARES}$ provided by measurements of the
ANTARES Collaboration and inferable standard deviation $\sigma_\epsilon$.
Deviations need to be allowed because missing optical modules lead to wrong
assumptions about the efficiencies in one storey (see \cite{antares:eff}).

In addition, the photon sensitive area of the optical modules is modeled as a
circular surface with an area of $A_\text{om} = \pi r_{\text{om}}^2$. Only a
fraction 
\begin{align} 
  \frac{A_\text{om}}{A_{\text{light}}(r_i(t))} 
\end{align} 
of the emitted photons can hit the detector $i$ since the photons spread
uniformly on the surface of the sphere $A_{\text{light}}(\vec{r}) = 4 \pi
\cdot r_i^2(t)$ at a distance $r_i(t) = |\vec{r}_i(t)|$ from the source. The
vector between the light source position $\vec{x}$ and the detector position
$\vec{p}_i$ is calculated by
\begin{align*} 
  \label{eq:relsourcepos}
  \vec{r}_i(t) = \vec{p}_{i} - \vec{x}(t).  
\end{align*} 

Furthermore, the angular acceptance and accordingly the orientation of the
detector plays an important role. The angular acceptance
$\alpha(\theta_\text{optical})$ as a function of the optical angle
$\theta_\text{optical}$ is provided by the ANTARES experiment. The cosine of
the optical angle depends on the orientation $\vec{o}_{i}$ of the optical module
and can be obtained by
\begin{align*} \cos{\theta_{\text{optical}}} =
  \frac{-1}{|\vec{r}_i(t)||\vec{o}_{i}|} \vec{r}_i(t) \cdot \vec{o}_{i}.  
\end{align*} 
The polynomial fit of the angular acceptance stated as
\begin{align} 
  \alpha(\theta_\text{optical}) = \alpha(\vec{r}_i(t), \vec{o}_{i}) 
\end{align} 
is used to calculate the percentage of photons coming from direction
$\vec{r}_i(t)$ that actually hit the optical module oriented in $\vec{o}_{i}$
direction.

Finally, the impact of electromagnetic absorption by water and the photon
scattering can be determined by the attenuation length $l_{\text{att}}$ and
the Beer-Lambert law. The fraction of photons reaching the detector after
absorption and scattering can be calculated by
\begin{align} 
  \frac{\widetilde{\mathcal{N}}_i(t)}{\mathcal{N}(t)} = e^{-r_i(t)/l_{\text{att}}}
\end{align}
with $\mathcal{N}(t)$ as defined before being the number of emitted photons and
$\widetilde{\mathcal{N}}_i(t)$ the number of photons reaching the distance of the
detector $i$. Despite small local changes of the water properties in the deep
sea the attenuation length $l_{\text{att}}$ is assumed to be independent of the
position of the optical modules. Due to the measurements of the attenuation
length mentioned in section~\ref{sec:data} (Table~\ref{tab:light_att}), which
vary over the years, we assume a flat prior distribution for $l_\text{att}$ within
the interval $l_\text{att} = (45 \text{ m}, 60 \text{ m})$. This assumption
prevents the reconstruction to adapt unphysical values and thereby stabilizes
the inference.

Combining these effects on the emitted photons, the response function for the
photon counts $\lambda_{i,t}$ detected by the optical module $i$ can be
expressed as
\begin{align} 
  \lambda_{i}(t) =
  \epsilon_{i} \cdot \mathcal{N}(t)  \cdot \alpha(\vec{o}_i,
  \vec{r}_i(t)) \cdot e^{-\frac{r_i(t)}{l_{\text{att}}}} \frac{A_{\text{om}}}{4 \pi \cdot
  r^2_i(t)}.  \label{eq:response} 
\end{align} 
It is important to highlight that $\mathcal{N}$ describes the number of
photons isotropically emitted by a hypothetical luminescent organism modeled
as point source. The complex structure of real biological organisms that may
lead to anisotropic emission is not covered by our model. This mismatch
between the assumed isotropic model and the real light emission of organisms
in the deep sea can lead to unrepresentative uncertainty estimates of the
MGVI algorithm, as discussed in section~\ref{sec:data}. A compact summary of
the response function with its high-level parameters is given in table~\ref{tab:response}. 
A visualization of the described generative model is
given in figure~\ref{fig:response_tree} at the end of the section.

\begin{table*} 
  \caption{Overview of the response function with its parameters and their
  explanations.} 
  \label{tab:response} 
  \begin{tabular}{c} 
    \\
    $\lambda_{i}(t) =
    \epsilon_{i} \cdot \mathcal{N}(t)  \cdot \alpha(\vec{o}_i,
    \vec{r}_i(t)) \cdot e^{-\frac{r_i(t)}{l_{\text{att}}}} \frac{A_{\text{om}}}{4 \pi \cdot
    r^2_i(t)}$\\
    \\
  \end{tabular} 
  \begin{tabular}{c  l} 
    \hline
    High-level parameters & Explanation \\ 
    \hline
    $\epsilon_{i}$ & Detector efficiency of optical module $i$, assumed constant over time\\ 
    $\mathcal{N}(t)$ & Emitted photons of an isotropic point source emitter,
    time dependent \\ 
    & \\ 
    $\vec{x}(t)$ & Position of biological object, time dependent\\ $\vec{p}_{i}$ &
    position of optical module $i$, assumed constant over time\\ 
    $\vec{r}_i(t) = \vec{p}_{i}-\vec{x}(t)$ & Vector from source to optical
    module $i$\\ 
    $r_i(t) = | \vec{r}_i(t) |$ & Distance from source to optical module $i$\\ 
    $\vec{o}_i$ & Orientation of optical module $i$\\ 
    $\alpha(\vec{o}_i, \vec{r}_i(t))$ & Angular acceptance\\ 
    & \\ 
    $l_{\text{att}}$ & Attenuation length of light in sea water, assumed constant over time \\ 
    & \\ 
  $A_\text{om} = \pi r_{\text{om}}^2$ & Effective area of the optical module assuming a
    circle\\ 
    $A_{\text{light}} = 4 \pi \cdot r_i^2(t)$ & Area covered by spherical
    radiation at the location of optical module $i$\\ 
    \hline \\ 
  \end{tabular}
\end{table*}

\subsubsection{Luminosity}
\label{sec:lumin}

The bioluminescence lightcurves are assumed to be the dominant feature of the
photon counts over time. The luminosity model $\mathcal{L}(t) =
\frac{\mathcal{N}(t)}{\Delta t}$ has to be able to capture all features of a
bioluminescence flash and hence be able to provide sensible prior samples.

Since the number of emitted photons is always positive, $\mathcal{N}(t)$ can
be described to sufficient accuracy by a log-normal model. The lightcurve
structure does not depend on absolute times $t$, but on the relative timing
$\Delta t = t_l - t_k$. Therefore, a correlated signal $s(t)$ with a given
correlation matrix $C_{t_l \, t_k} = C(t_l - t_k)$ is used to model the burst
kinetics under the assumption of statistical homogeneity. Combining all prior
assumptions the equation for the luminosity model yields
\begin{align}
  \mathcal{N}(t) = e^{s(t)}, 
\end{align} 
with $s(t)$ sampled from a Gaussian with inferable mean $\mu_s$ and fixed
covariance $C$,
\begin{align} 
  s(t) \hookleftarrow \mathcal{G}(s-\mu_s, C).  
\end{align}
Although we do not know the underlying correlation structure exactly, we can
use the recorded lightcurve structure to estimate the correlation. The main
motivation for a fixed covariance is to reduce the computation time.
Alternatively, the correlation could be infered as well, as it is done for
the source movement in section~\ref{sec:movement}.
The formulation and discussion of a reasonable correlation function as well as
the distribution transformations used for the luminosity model can be found in
the \hyperref[appendix:distr_trans]{appendix}.  The
parameters of the luminosity model were chosen such that the attributes of
recorded bursts given in table~\ref{tab:bio_attr} were fulfilled and similar
bursts as in figure~\ref{fig:example_data} could be constructed.  A compact
summary of the function modeling the photon rate with its parameters is given in 
table~\ref{tab:luminosity} and illustrated in figure~\ref{fig:response_tree}.\\ 

\begin{table}[!h] 
  \caption{Overview of the emitted photon number function with its parameters,
  their explanations and origin.} 
  \label{tab:luminosity} 
  \begin{tabular}{c} 
    \\
    $\mathcal{N}(t) = e^{s(t)}$\\
    \\
  \end{tabular}\\
  \begin{tabular}{c l l} 
    \hline
    Parameters & Explanation & Origin \\ 
    \hline
    $s$ & Correlated field & $s \hookleftarrow \mathcal{G}(s-\mu_s, C)$ \\ 
    $\mu_s$ & Inferable mean & See \hyperref[appendix:corr_matrix]{appendix}\\
    $C$ & Correlation matrix & Fixed prior value \\
    \hline
  \end{tabular}
\end{table}

\subsubsection{Source movement}
\label{sec:movement}

The $x$, $y$ and $z$ directions of the movement of the source within the
coordinate system of the detector are handled independently from each other.
For each direction $j \in {x,y,z}$ a velocity vector $v_j(t)$ can be reconstructed
to describe the movement starting at point $\vec{x}_0 = (x_0, y_0, z_0)$.

Similarly to the luminosity model, statistical stationarity is assumed for
the velocity as a function of time. But instead of using a fixed correlation
matrix, this is inferred as well. In contrast to the luminosity model, we do
not have access to previous recordings of the movement to estimate the
correlation. Consequently, the velocity vectors
\begin{align} 
  v_j(t) \hookleftarrow \mathcal{G}(v_j - \mu_{v_j}, K)
\end{align} 
are sampled from a Gaussian with an inferable correlation matrix
$K_{t_l t_k}$ as covariance. The covariance can be set such that sampled
velocities meet criteria of biological plausible movements.
A detailed discussion about reconstructing correlation
functions of a signal can be found in the NIFTy documentation
\cite{ift:niftysoftware} and the corresponding papers \cite{ift:nifty5, ift:M87}.

The starting position of the movement $\vec{x}_0$ is assumed to be drawn from
a uniform prior distribution within a box $j_0 = (j_0^-, j_0^+)$ for each
direction $j \in {x,y,z}$,
\begin{align}
  \label{eq:uni_position}
  j_0 \hookleftarrow \mathcal{U}(j_0^-, j_0^+).
\end{align}
A reasonable limitation of the source location is possible due to the setup of
the ANTARES telescope and the assumption of isotropic light emissions. By
detecting the optical module with the highest photon numbers during a
bioluminescence flash, source locations outside of a given volume around this module
can be excluded.

The absolute position $\vec{x}(t)$ at time $t$ can be obtained by integrating
the velocity from start time $t_0$ up to $t$ and adding the start position
$\vec{x}_0$. Hence, the expression of the position vector $\vec{x}(t)$ can be
derived as 
\begin{align} \vec{x}(t) = \vec{x}_0 + \Delta t
  \cdot \sum_{t_k=t_0}^t \vec{v}(t_k).
\end{align} 

\begin{figure*}[p]
  %\resizebox{0.95\textwidth}{!}{%
  \includegraphics[width=0.95\textwidth, keepaspectratio]{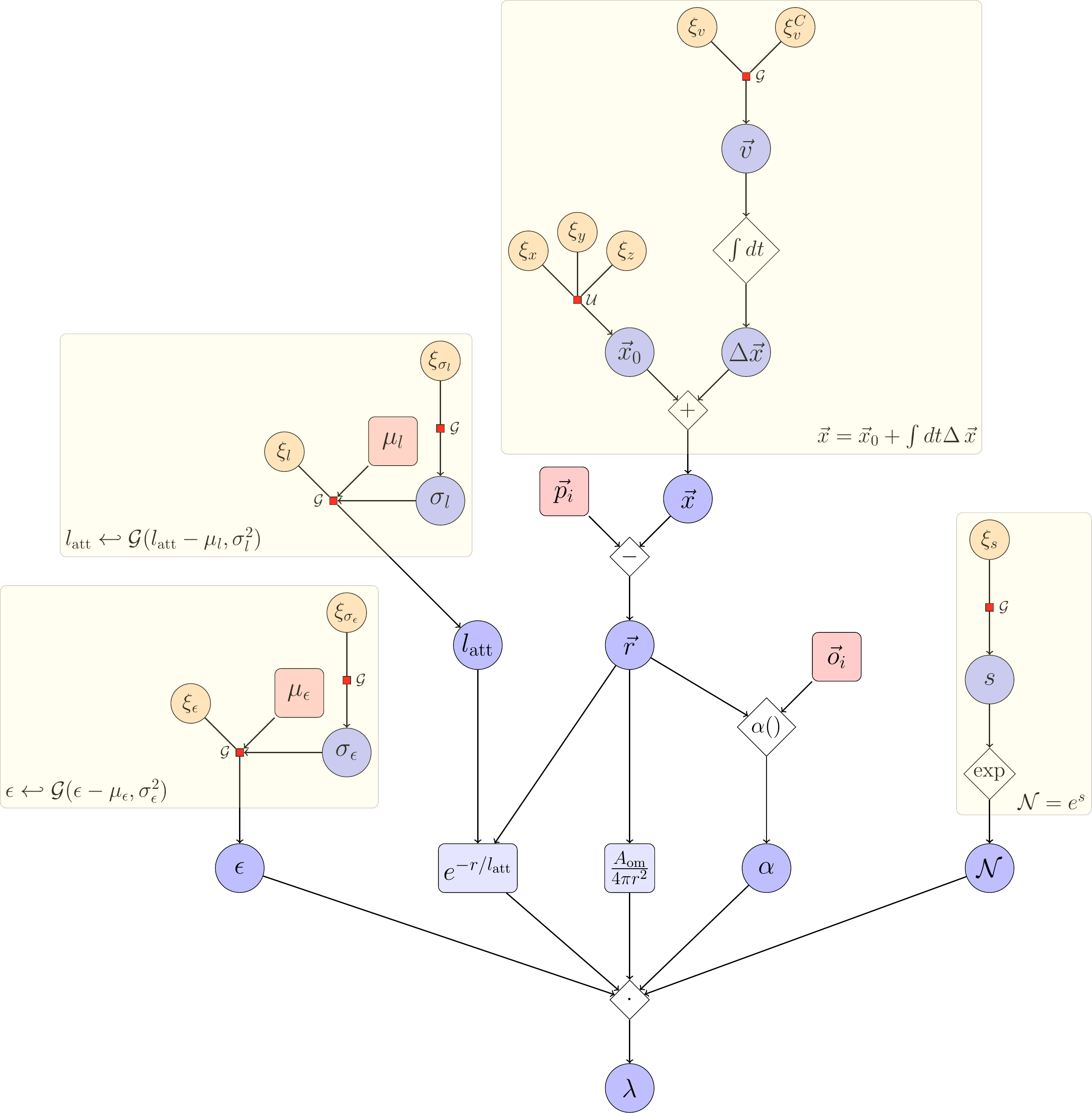}
  \caption{Generative model of the response function. Blue shapes indicate in
  quantities that are in principle observable. The standardized variables are colored
  orange and operations have a diamond shape. The transformations from a
  standard Gaussian distribution into a target distribution are labeled as a
  small red square with the target distribution next to it. All red colored
  values include a priori assumptions, i.e. distribution transformation,
  detector position and orientation.
}  
  \label{fig:response_tree} 
\end{figure*}

Distribution transformations between Gaussian distributions and uniform
distributions $x_0(\xi_{x_0})$ are used and discussed in the
\hyperref[appendix:distr_trans]{appendix}. A compact summary of the position and movement
model with its parameters is given in figure~\ref{fig:response_tree} and in table
\ref{tab:position}. 

\begin{table}[!h] 
  \caption{Overview of the position and movement model with its parameters,
  their explanations and origin.}
  \label{tab:position} 
  \begin{tabular}{c} 
    \\
    $\vec{x}(t) = \vec{x}_0 + \Delta t \cdot \sum\limits_{t_k=t_0}^t \vec{v}(t_k)$\\ 
    \\
  \end{tabular}
  \begin{tabular}{c l l} 
    \hline
    Parameters & Explanation & Origin \\ 
    \hline
    $\vec{x}_0$ & Starting position & $j_0 \hookleftarrow \mathcal{U}(j_0^-,j_0^+)$\\ 
    & $x,y,z$ independent & \\ 
    $(j_0^-, j_0^+)$ & Range for & Fixed prior values \\ 
    & uniform distribution & \\ 
    & & \\ 
    $\vec{v}$ & Velocity vector & $v_j \hookleftarrow \mathcal{G}(v_j -
    \mu_{v_j}, K)$\\ 
    & $v_x,v_y,v_z$ independent & \\ 
    $\mu_{v_j}$ & Inferable mean & See \cite{ift:M87}\\
    $K$ & Inferable & \\
    & correlation matrix & A priori assumptions
    \\
    \hline
  \end{tabular} 
\end{table}

\subsubsection{Optical background}
\label{sec:background}

In addition to the lightcurves emitted by an individual luminescent organism,
photons from other sources are detected as well. As mentioned in section
\ref{sec:data} an almost constant background light is assumed to be induced by
nuclear decays and bacterial bioluminescence. We model this optical background
as constant offset $n_\text{storey}$ for each individual optical module.
Furthermore, we assume that all optical modules of one storey monitor a water
volume containing a similar amount of nuclear decays and of small luminescent
organisms. Hence, the same constant light background is recorded at one storey.
Allowing local variations of the water attributes, each background value
$n_\text{storey}$ for one storey  is sampled from a Gaussian distribution

\begin{align}
  \label{eq:noise}
  n_\text{storey} \hookleftarrow \mathcal{G}(n_\text{storey} - \mu_n,
\sigma_n^2).
\end{align}
This leads to the extended response function of an optical module
in a given storey
\begin{align}
  \lambda_{i, \text{ext}} = \lambda_i + \epsilon_i \cdot n_\text{storey}.
  \label{eq:ext_response}
\end{align}

\subsubsection{Reconstruction routine}
\label{sec:reconstruction_routine}

For the inference of such a complex model we propose a reconstruction routine
that is split up into two parts. We first assume a simpler model by
neglecting the movement of the source. This assumption reduces the complexity
of the model and allows a more stable inference. After analyzing the results
of this static reconstruction, one can conclude whether a more complex model
including the movement of the source is required. If the data can be
reconstructed by the assumption of a static source, the real source does not
move or its movement can not be resolved. Observing deviations between the
data and its reconstruction imply the need for a better model. In that case
the reconstructed position and luminosity are used as initial samples for the
dynamic reconstruction. The samples of remaining variables of the model are
randomly initiated according to their model priors. This scheme reduces the
risk of overfitting to which such a complex model is prone to. In the next
section the splitting of the reconstruction is performed on simulated data to
analyze this routine, as well as the limitations and opportunities of the
model.

%%%%%%%%%%%%%%%%%%%%%%%%%%%%%%%%%%%%%%%%%%%%%%%%%%%%%%%%%%%%%%%%%%%%%%%%%%%%%%%
% SIMULATION
%%%%%%%%%%%%%%%%%%%%%%%%%%%%%%%%%%%%%%%%%%%%%%%%%%%%%%%%%%%%%%%%%%%%%%%%%%%%%%%

\section{Simulations}
\label{sec:simulation}

In the last section we developed the response function of the expected photon
counts of an optical module. Due to the form of the response, being a product 
of the factors $\mathcal{N}$, $\epsilon$, $e^{r/l_\text{att}}$
and $\frac{A_\text{om}}{4 \pi r^2} \cdot \alpha(\vec{r},\vec{o})$,
increases of one factor can be compensated by decrements of another factor,
and vice versa. This leads to a degeneracy between the emitted photon number
$\mathcal{N}$, the source position $\vec{x}$, the efficiency $\epsilon$ of an
optical module and the attenuation length $l_\text{att}$. This degeneracy
can only be reduced by using the data of multiple OMs and assuming a constant
background light. In the following sections we analyze the degeneracy between
those variables by performing the reconstruction on synthetic data sets of
multiple OMs. First, we focus on a static reconstruction of a simulated
static source in order to examine whether the degeneracy can be reduced to
such a degree that a light source can be localized and the corresponding
bioluminescence lightcurve reconstructed. Secondly, we apply the complete
dynamic reconstruction to the simulated static source, as well as a simulated
dynamic source. These results are used to discuss the spatial resolution of
the reconstruction and the possibility of a movement reconstruction.

For the simulation of static and dynamic sources, the configuration of the
ANTARES telescope from some past moment in time is used to create the detector
setup. The efficiencies are randomly drawn from Gaussian distributions
(equation~\ref{eq:efficiency}) with fixed standard deviation
for each optical module. To simulate a realistic environment, the
attenuation length is set to $l_\text{att} = 55$~m and the constant light
background of each storey is drawn from a Gaussian distribution 
(equation~\ref{eq:noise}) with fixed mean $\mu_n = 50~\text{ kHz}$ and fixed standard
deviation $\sigma_n = 5 \text{ kHz}$. Further, the shape of a burst is
extracted from real data and scaled to reach realistic luminosity values.

The initial position in both cases is manually set to ($x_0=50$~m, $y_0=1$~m,
$z_0=23$~m). While the static source stays at its initial position, we simulate
a linear movement with a velocity $v= 0.2~\frac{\text{m}}{\text{s}}$ for the
dynamic source. The light signal is detected by $15$ optical modules
distributed over $5$ storeys of one line. The mean of the recorded photon
hits of each optical module is calculated according to equation~\ref{eq:ext_response}. 
Tables~\ref{tab:recon_static_source_obs} and
\ref{tab:recon_dynamic_static_source_obs} provide the ground truth
observables of the static and dynamic simulated source, respectively.

\subsection{Static reconstruction of a synthetic static source}
\label{sec:static_static_source}

The parameters for the initial position of the static inference model
(equation~\ref{eq:uni_position}) used to reconstruct data generated from the
static source are provided in table~\ref{tab:static_source_param}. The
parameters of the photon number model are given in the
\hyperref[appendix:corr_matrix]{appendix}. Assumptions about the efficiencies of the optical
modules, the attenuation length and the light background are given in table~\ref{tab:static_det_param}.
The position and orientation of each optical
module used for the simulation are also used for the inference.

\begin{table}
  \caption{Model parameters of a static inference source. The model is used to reconstruct
  the photon count data generated from a simulated static source.} 
  \label{tab:static_source_param} 
  \begin{tabular}{l l l} 
    \hline
    Observable & Model & Model parameters \\
    \hline
    $x$ position & Uniform & $x =(15\text{ m}, 75\text{ m})$\\
    $y$ position & Uniform & $y=(-20\text{ m}, 40\text{ m})$\\
    $z$ position & Uniform & $z=(15\text{ m}, 35\text{ m})$\\ 
    Flash shape & Correlated signal & See \hyperref[appendix:corr_matrix]{appendix}\\
    \hline
  \end{tabular}
\end{table}

\begin{table}
  \caption{Model parameters used to reconstruct the attributes of a
  simulated telescope and its surroundings.} 
  \label{tab:static_det_param} 
  \begin{tabular}{l l l} 
    \hline
    Observable & Model & Model parameter \\
    \hline
    $\epsilon$ & $\mathcal{G}(\epsilon - \mu_\epsilon, \sigma_\epsilon^2)$  &
    $\mu_\epsilon = \epsilon_\text{ANTARES}$  \\ 
    & with $\mathcal{G}(\sigma_\epsilon - \mu_{\sigma_\epsilon},
    \sigma^2_{\sigma_\epsilon})$ & 
    $\mu_{\sigma_\epsilon} = 0.01$, \\ 
    & & $\sigma_{\sigma_\epsilon} = 0.001$ \\ 
    && \\
    $l_\text{att}$ & $\mathcal{U}(l_\text{min}, l_\text{max})$ & $l_\text{min} = 45 \text{
    m}$, \\ 
    & & $l_\text{max} = 60 \text{ m}$ \\ 
    && \\
    $n_\text{storey}$ & $\mathcal{G}(n_\text{storey} - \mu_n, \sigma_n^2)$ &
    $\mu_n = 50 \text{ kHz}$, \\
    & & $\sigma_n = 5 \text{ kHz}$ \\
    \hline
  \end{tabular}
\end{table}

Several reconstruction runs were performed, each with a different random seed
to assess numerical stability. In order to analyze the degeneracy between
source position, luminosity and attenuation length, we compare the
reconstructed values with their ground truth summarized in table~\ref{tab:recon_static_source_obs}. 
The position of the light source $\vec{x} = (50 \text{ m}, 1 \text{ m}, 23 \text{
m})$ could be reconstructed for each run. The ground truth position lies
within the given error ranges, which does not include systematic errors. In
contrast, for each reconstruction run a different attenuation length was
inferred and the reconstructed peak luminosity deviates up to $5 \%$ from the
ground truth.

We conclude that the degeneracy between the luminosity and the attenuation
length, i.e. the fact that any decrease in observed photon counts could
either be explained by a shorter attenuation length or a less bright
source, still remains for the given data and used model. In future work, a
better informed model might help to break the degeneracy. However, this
degeneracy has negligible impact on the localization of the source. The ratio
of photons counts between the optical modules $\frac{\lambda_i}{\lambda_j}$,
which is independent of the luminosity and attenuation length, provides
information about the source position and hence makes it unambiguously
inferrable.

\begin{table*}
  \caption{Ground truth and reconstructed parameters for a static inference
  source. The synthetic photon count data were generated from the simulated
  static source. The reconstructed attenuation length is
  also included.}
  \label{tab:recon_static_source_obs} 
  \begin{tabular}{l r r r r} 
    \hline
    Observable & Ground truth & Run 1 & Run 2 & Run 3\\
    \hline
    $x$ source position (m) & $50.00$ & $49.96 \pm 0.04$ & $49.94 \pm 0.04$ & $49.96 \pm 0.06$\\
    $y$ source position (m) & $1.00$ & $0.96 \pm 0.07$ & $1.03 \pm 0.05$ & $0.95 \pm 0.08$\\
    $z$ source position (m) & $23.00$ & $23.03 \pm 0.03$ & $22.97 \pm 0.03$ & $23.03 \pm 0.03$\\
    \\
    Flash duration (s) & $5.14$ & $5.14$ & $5.14$ & $5.14$\\
    $\mathcal{L}_\text{max}$ ($10^{10}$ Hz) & $3.71$ & $3.66 \pm 0.07$ & $3.51 \pm 0.02$ & $3.65 \pm 0.01$\\
    Total emission ($10^{10}$ photons flash$^{-1}$) & $5.20$ & $5.14$ & $4.94$ & $5.13$\\
    \\
    Attenuation length (m) & $55.00$ & $53.90 \pm 2.09$ & $59.94 \pm 0.20$ & $54.23 \pm 3.06$\\
    \hline
  \end{tabular}
\end{table*}

\begin{table*}
  \caption{Ground truth and reconstructed parameters using a dynamic source model. The
  synthetic photon count data were generated from a simulated static source.
  The positions are given at the beginning of the observation period $t_0$, during
  the time of highest photon emission $t_\text{h}$ and at the end of the data set ($t_\text{max}$) to
  reflect the movement of the source. 
}
  \label{tab:recon_static_dynamic_source_obs} 
  \begin{tabular}{l r r r r} 
    \hline
    Observable & Ground truth & Run 1 & Run 2 & Run 3 \\
    \hline
    $x(t_0)$ source position (m) & $50.00$ & $49.89 \pm 0.08$ & $49.88 \pm 0.09$ & $49.89 \pm 0.08$ \\
    $x(t_\text{h})$ source position (m) & $50.00$ & $49.96 \pm 0.05$ & $49.96 \pm 0.05$ & $49.96 \pm 0.05$ \\
    $x(t_\text{max})$ source position (m) & $50.00$ & $50.02 \pm 0.08$ & $50.03 \pm 0.09$ & $50.03 \pm 0.08$ \\
    $y(t_0)$ source position (m) & $1.00$ & $0.92 \pm 0.06$ & $0.91 \pm 0.06$ & $0.95 \pm 0.06$ \\
    $y(t_\text{h})$ source position (m) & $1.00$ & $1.00 \pm 0.07$ & $1.00 \pm 0.07$ & $1.00 \pm 0.06$ \\
    $y(t_\text{max})$ source position (m) & $1.00$ & $1.07 \pm 0.12$ & $1.07 \pm 0.13$ & $1.04 \pm 0.09$ \\
    $z(t_0)$ source position (m) & $23.00$ & $23.04 \pm 0.08$ & $23.02 \pm 0.06$ & $23.02 \pm 0.07$ \\
    $z(t_\text{h})$ source position (m) & $23.00$ & $23.00 \pm 0.04$ & $23.00 \pm 0.04$ & $23.00 \pm 0.04$ \\
    $z(t_\text{max})$ source position (m) & $23.00$ & $22.96 \pm 0.10$ & $22.99 \pm 0.08$ & $22.98 \pm 0.09$ \\
    \\
    Flash duration (s) & $5.14$ & $5.14$ & $5.14$ & $5.14$ \\
    $\mathcal{L}_\text{max}$ ($10^{10}$ Hz) & $3.71$ & $3.51 \pm 0.02$ & $3.51 \pm 0.02$ & $3.51 \pm 0.02$ \\
    Total emission ($10^{10}$ photons flash$^{-1}$) & $5.20$ & $4.92$ & $4.93$ & $4.93$ \\
    \hline
  \end{tabular}
\end{table*}

\subsection{Dynamic reconstruction of a synthetic static source}
\label{sec:static_dynamic_source}

As explained in section~\ref{sec:reconstruction_routine} the flash lightcurve
and the position of the static reconstruction of the previous section are
used as the initially assumed position for the dynamic reconstruction. In
table~\ref{tab:static_dynamic_source_param} the parameters of the dynamic
model are summarized. The model parameters of the detector and its
environment are the same as in the static reconstruction and provided in
table~\ref{tab:static_det_param}.

Since the reconstructions of the remaining observables were already discussed
in the previous section and similar results could be observed for assuming a
dynamic model, we focus on the discussion about the velocity. For all runs, a
source movement was reconstructed with a mean velocity $v \simeq 0.01
\frac{\text{m}}{\text{s}}$. The results of the different reconstruction runs
with varying seeds are summarized in table~\ref{tab:recon_static_dynamic_source_obs}.

These results show that a source movement with a mean velocity $v = 0.01
\frac{m}{s}$ can not be distinguished from a static source for this detector
setup. This automatically defines a lower bound for the spatial resolution
of this method using the ANTARES detector. The spatial resolution depends on 
the setup of the detector, i.e. number of optical modules and angular acceptance.
Therefore, future neutrino telescopes with different detector setups might increase the resolution. 
In the next section we demonstrate that a movement reconstruction
is theoretically possible with the ANTARES setup by applying the method to
synthetic data drawn from a simulated dynamic source.

\begin{table}
  \caption{Model parameters of a dynamic inference source. The model is used
  to reconstruct the photon count data generated from a simulated static
  source.}
  \label{tab:static_dynamic_source_param} 
  \begin{tabular}{l l l} 
    \hline
    Observable & Model & Model parameters \\
    \hline
    $v_i$ velocity  & Correlated signal & $\mu_v = 0 \frac{\text{m}}{\text{s}}, 
    \sigma_v = 0.1 \frac{\text{m}}{\text{s}},$\\
    & with $i \in \{x,y,z\}$  & $\sigma_v = 0.1 \frac{\text{m}}{\text{s}}$\\
    & & \\
    $x$ position & Uniform & $x =(15\text{ m}, 75\text{ m})$\\
    $y$ position & Uniform & $y=(-20\text{ m}, 40\text{ m})$\\
    $z$ position & Uniform & $z=(15\text{ m}, 35\text{ m})$\\ 
    & & \\
    Flash shape & Correlated signal & See \hyperref[appendix:corr_matrix]{appendix}\\
    \hline
  \end{tabular}
\end{table}

\subsection{Complete reconstructions of a dynamic source}
\label{sec:dynamic_source}

\subsubsection{Static reconstruction}
\label{sec:dynamic_static_source}

For the first part of the reconstruction routine, the same model parameters of
the detector and source are used as in the static scenario given in table
\ref{tab:static_source_param} and \ref{tab:static_det_param}. The
results of the reconstruction of a dynamic source using the static model are
provided in table \ref{tab:recon_dynamic_static_source_obs}.

\begin{table*}
  \caption{Ground truth and reconstructed parameters using a static source model. The
  synthetic photon count data were generated from a simulated dynamic source.
  The summary includes the reconstructed attenuation length.}
  \label{tab:recon_dynamic_static_source_obs}
  \begin{tabular}{l r r r r} 
    \hline
    Observable & Ground truth & Run 1 & Run 2 & Run 3\\
    \hline
    $x$ source position (m) & $50.02 \rightarrow 51.15$ $\rightarrow$ $52.10$ & $51.64 \pm 0.04$ & $50.80 \pm 0.03$ & $51.14 \pm 0.05$\\
    $y$ source position (m) & $1.02 \rightarrow \ \: 2.15 \rightarrow \ \: 3.10$ & $1.79 \pm 0.06$ & $2.65 \pm 0.04$ & $2.25 \pm 0.05$\\
    $z$ source position (m) & $23.02 \rightarrow 24.15$ $\rightarrow$ $25.10$ & $23.98 \pm 0.03$ & $24.55 \pm 0.03$ & $24.30 \pm 0.02$\\
    \\
    Flash duration (s) & $5.14$ & $5.14$ & $5.14$ & $5.14$\\
    $\mathcal{L}_\text{max}$ ($10^{10}$ Hz) & $3.71$ & $3.49 \pm 0.02$ & $3.28 \pm 0.02$ & $3.99 \pm 0.03$\\
    Total emission ($10^{10}$ photons flash$^{-1}$) & $5.20$ & $4.94$ & $4.65$ & $5.65$\\
    \\
    Attenuation length (m) & $55.00$ & $59.98 \pm 0.06$ & $59.92 \pm 0.24$ & $45.12 \pm 0.45$\\
    \hline
  \end{tabular}
\end{table*}

\begin{figure}

  \includegraphics[width=0.45\textwidth]{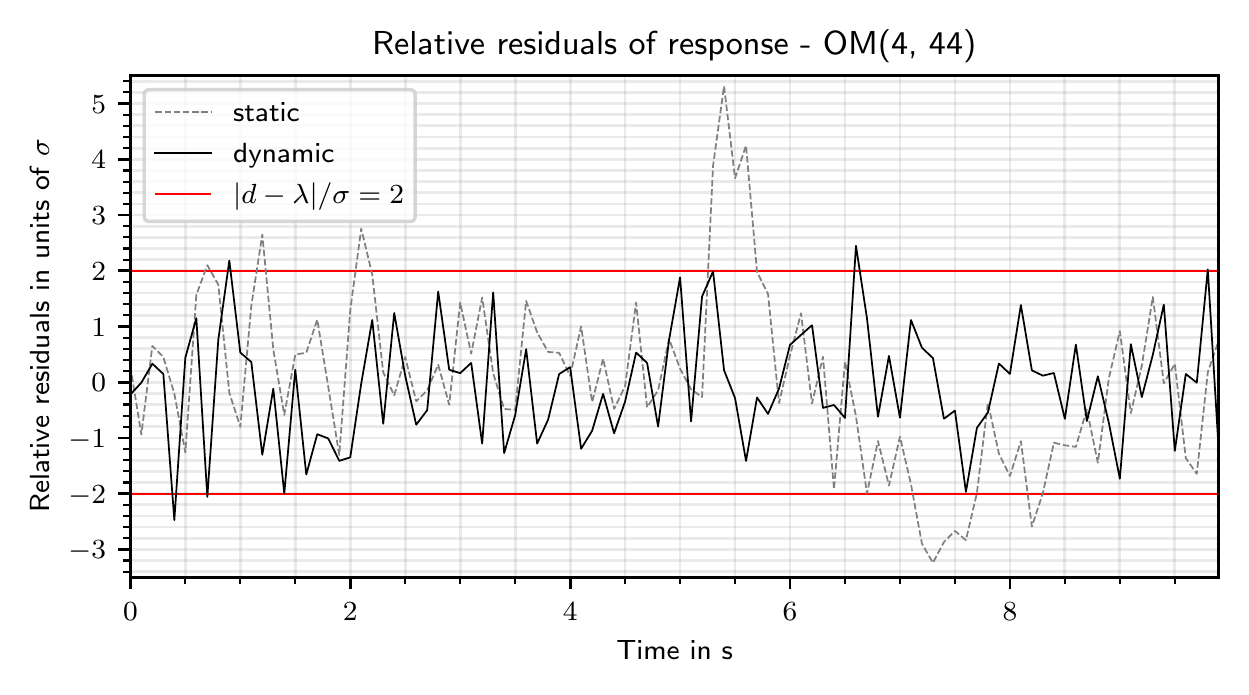}
  \caption{
   Comparison of relative residuals of a static (dashed, gray) and dynamic
   (solid, black) inference source. The synthetic data were generated from a
   simulated dynamic source. The red line marks deviations as large as twice the shot
   noise $\sigma$ of a Poisson process.
  }
  \label{fig:recon_response_dynamic_dynamic_44} 
\end{figure}

Analyzing the results shows that a dynamic reconstruction might increase the
accuracy of the estimates as the static model did not provide an optimal
fit. In figure~\ref{fig:recon_response_dynamic_dynamic_44} the residuals $e
= \frac{d - \lambda}{\sigma}$ relative to the shot noise of the Poissonian
measurement process $\sigma = \sqrt{\lambda}$ of optical module $(4, 44)$ are
presented. During the light flash an increased level of deviations ($2 < |e|
< 5$) can be recognized. Residuals of optical module $(4,44)$ close to zero
can be found at $t_\text{min} \simeq 6.3$~s. Detailed analysis shows that
the relative residuals are positive $e > 0$ before the minimum $t <
t_\text{min}$ and negative $e < 0$ afterwards $t > t_\text{min}$. Even though
these variations are only slightly above the shot noise of the measurement
process $\sigma$, they can be explained by a moving source. Hence, another
reconstruction is performed to reduce these residuals by introducing a model
for the source movement.

\subsubsection{Dynamic reconstruction}
\label{sec:dynamic_dynamic_model}

We use the same dynamic model as in the previous section
\ref{sec:dynamic_static_source}. The parameters used to define the source
model are given in table \ref{tab:static_dynamic_source_param} and the
parameters for the detector setup and its environment are presented in table
\ref{tab:static_det_param}.

\begin{figure*}[p]
  \includegraphics[width=0.9\textwidth]{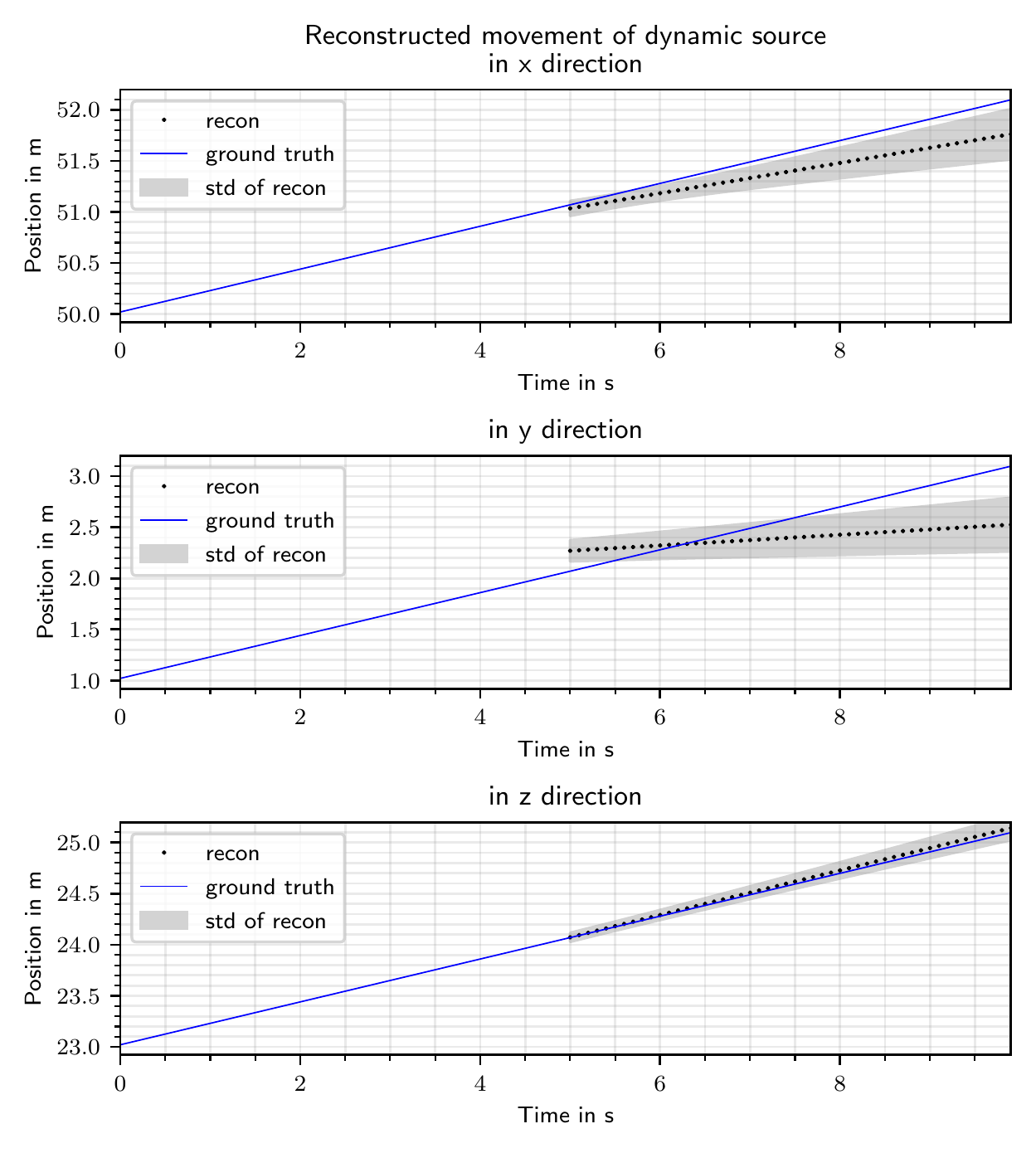}
  \caption{Reconstruction of the source position with the dynamic model. The
    simulated position (solid, blue) is shown in comparison to the reconstructed
    (circles, black) position. The reconstruction is only given for the time
    duration of the burst. Shown is also the uncertainty of the position
    reconstruction, shaded in gray.
  } 
  \label{fig:recon_pos_dynamic_1D} 
\end{figure*}

The increased level of variations between simulated and reconstructed response
of the static model can be almost reduced to $0 < |e| < 2$ by the dynamic
reconstruction as shown in figure
\ref{fig:recon_response_dynamic_dynamic_44}, a clear improvment over the
static model. The reconstructed movement visualized in figure
\ref{fig:recon_pos_dynamic_1D} highlights the limits of the ANTARES detector
for tracking individual organisms. A reasonable estimate of the movement
could be reconstructed for a simulated light source moving linearly with a
velocity $v = 0.2 \frac{\text{m}}{\text{s}}$. Although the reconstructed
locations in $y$ dimension display deviations from the ground truth, the
residuals are close to the shot noise of the measurement process. These
results show that a linear movement with a velocity $v = 0.2
\frac{\text{m}}{\text{s}}$ is in priciple resolvable, but is not free of
systematics. Besides the degeneracies, which are already discussed in section~\ref{sec:static_static_source},
the angular acceptance has great impact on the
positioning of a light source. The optical modules of the ANTARES detector
have a wide angular acceptance \cite{antares:opticalmodule}. Therefore,
changes of the source position in non-radial directions do not lead to
significant changes in the photon count numbers, which reduces the
possibility to recognise position changes. Multiple optical modules with
smaller angular acceptance might increase the spatial resolution. The
reconstructed flash characteristics differ only slightly between the
reconstruction with a static and a dynamic model as shown in table~\ref{tab:recon_dynamic_dynamic_source_obs}. Therefore, with regards to the
bioluminescence flash lightcurves, both models are sufficient to reconstruct
a reasonable estimate for velocities below $v = 0.2
\frac{\text{m}}{\text{s}}$.

\begin{table*}
  \caption{Ground truth and reconstructed parameters using a dynamic source
  model. The synthetic photon count data were generated from a simulated
  dynamic source. The positions are given at the beginning of the observation
  period $t_0$, during the time of highest photon emission $t_\text{h}$ and
  at the end of the data set ($t_\text{max}$) to reflect the movement of the
  source.
}
  \label{tab:recon_dynamic_dynamic_source_obs} 
  \begin{tabular}{l r r r r} 
    \hline
    Observable & Ground truth & Run 1 & Run 2 & Run 3 \\
    \hline
    $x(t_0)$ source position (m) & $50.02$ & $50.29 \pm 0.30$ & $50.02 \pm 0.13$ & $49.89 \pm 0.12$ \\
    $x(t_\text{h})$ source position (m) & $51.15$ & $51.09 \pm 0.07$ & $52.27 \pm 0.04$ & $52.23 \pm 0.04$ \\
    $x(t_\text{max})$ source position (m) & $52.10$ & $51.76 \pm 0.24$ & $54.15 \pm 0.10$ & $54.17 \pm 0.10$ \\
    $y(t_0)$ source position (m) & $1.02$ & $2.01 \pm 0.13$ & $1.04 \pm 0.13$ & $1.12 \pm 0.06$ \\
    $y(t_\text{h})$ source position (m) & $2.15$ & $2.29 \pm 0.11$ & $3.31 \pm 0.05$ & $3.33 \pm 0.05$ \\
    $y(t_\text{max})$ source position (m) & $3.10$ & $2.53 \pm 0.26$ & $5.20 \pm 0.12$ & $5.17 \pm 0.09$ \\
    $z(t_0)$ source position (m) & $23.02$ & $22.98 \pm 0.12$ & $22.92 \pm 0.08$ & $22.92 \pm 0.12$ \\
    $z(t_\text{h})$ source position (m) & $24.15$ & $24.16 \pm 0.04$ & $25.27 \pm 0.04$ & $25.21 \pm 0.04$ \\
    $z(t_\text{max})$ source position (m) & $25.10$ & $25.14 \pm 0.12$ & $27.22 \pm 0.10$ & $27.30 \pm 0.12$ \\
    \\
    Flash duration (s) & $5.14$ & $5.14$ & $5.14$ & $5.14$ \\
    $\mathcal{L}_\text{max}$ ($10^{10}$ Hz) & $3.71$ & $3.55 \pm 0.02$ & $3.61 \pm 0.03$ & $3.59 \pm 0.03$ \\
    Total emission ($10^{10}$ photons flash$^{-1}$) & $5.20$ & $5.14$ & $5.06$ & $5.03$ \\
    \hline
  \end{tabular}
\end{table*}

%%%%%%%%%%%%%%%%%%%%%%%%%%%%%%%%%%%%%%%%%%%%%%%%%%%%%%%%%%%%%%%%%%%%%%%%%%%%%%%
% DATA ANALYSIS
%%%%%%%%%%%%%%%%%%%%%%%%%%%%%%%%%%%%%%%%%%%%%%%%%%%%%%%%%%%%%%%%%%%%%%%%%%%%%%%
\section{Data analysis}
\label{sec:analysis}

After presenting the possibilities and limitations of the method, the
reconstruction routine is applied on data sets of the ANTARES Collaboration
using different random seeds. The complete routine consists of a
reconstruction using a static model and a dynamic model as explained in
section~\ref{sec:reconstruction_routine}. The outcomes of the static and
dynamic reconstruction provide similar results regarding the position during
the time of highest photon emission and the flash lightcurve. Therefore, it
is sufficient to present only the results of the final step using the dynamic
model.

For the reconstruction we rely on flash lightcurves that were detected over
several storeys to be able to reduce the degeneracy. Since the process of
finding such light emissions has not been automated, only a small excerpt of
flash patterns found in the ANTARES data is analyzed here. We identified
three suitable bioluminescence events in the data of early 2010, which we
analyse in the following. The data samples cover observation times of $7 -
10$ s. In table~\ref{tab:antares_flashes} we label the samples used for the
following reconstructions and state the array of optical modules that
detected the flash neglecting malfunctioning modules.

\begin{table}
  \caption{Data samples recorded by the ANTARES telescope reconstructed
  within this work.}
  \label{tab:antares_flashes} 
  \begin{tabular}{l c c c} 
    \hline
    Label & Time stamp (UTC) & OMs & Duration \\
    \hline
    Flash $1$ & 11.1.2010, 04:12:35 & $(4, 34 - 51)$ & $\sim 10$ s \\
    Flash $2$ & 11.1.2010, 04:13:20 & $(4, 37 - 54)$ & $\sim 9$ s \\
    Flash $3$ & 19.1.2010, 22:28:10 & $(4, 34 - 51)$ & $\sim 7$ s \\
    \hline
  \end{tabular}
\end{table}

\subsection{ANTARES recordings of flash $1$}

Starting with the 11th of January 2010 at 04:12:35 (UTC) a flash pattern was recorded
by 16 optical modules over 5 storeys. The complete recording 
has already been shown in figure~\ref{fig:example_data} in section~\ref{sec:data}.
Flash $1$ peaked at around $40$~s after the start of the recording.
Two optical modules, (4, 41) and (4,
51), within the optical module array (4, 37 - 51) did not record any photon
counts. The model parameters used for the reconstruction are presented in
table~\ref{tab:recon_420_param}.

\begin{table}[!h]
  \caption{Model parameters for the dynamic source model, the ANTARES telescope and
  its surroundings on the 11th of January 2010 at 04:12 (UTC). This model is used to
  reconstruct biological sources detected by the ANTARES telescope on 11th of
  January 2010 at 04:12 (UTC) and 04:13 (UTC), that emitted flash $1$ and $2$.}
  \label{tab:recon_420_param} 
  \begin{tabular}{l l l} 
    \hline
    Observable & Model & Model parameter \\
    \hline
    $\epsilon$ & $\mathcal{G}(\epsilon - \mu_\epsilon, \sigma_\epsilon^2)$  &
    $\mu_\epsilon = \epsilon_\text{ANTARES}$  \\ 
    & with $\mathcal{G}(\sigma_\epsilon - \mu_{\sigma_\epsilon},
    \sigma^2_{\sigma_\epsilon})$ & $\mu_{\sigma_\epsilon} = 0.05$,\\ 
    &&  $\sigma_{\sigma_\epsilon} = 0.01$ \\
    && \\
    $l_\text{att}$ & $\mathcal{U}(l_\text{min}, l_\text{max})$ &
    $l_\text{min} = 45 \text{ m}$,\\
    & & $l_\text{max} = 60 \text{ m}$
    \\ 
    && \\
    $n_\text{storey}$ & $\mathcal{G}(n_\text{storey} - \mu_n, \sigma_n^2)$ &
    $\mu_n = 50 \text{ kHz}$, \\
    & & $\sigma_n = 5 \text{ kHz}$ \\
    $x$ position & Uniform & $x =(15\text{ m}, 75\text{ m})$\\
    $y$ position & Uniform & $y=(-20\text{ m}, 30\text{ m})$\\
    $z$ position & Uniform & $z=(15\text{ m}, 35\text{ m})$\\ 
    $v_i$ velocity  & Correlated signal & $\mu_v = 0, \sigma_v = 0.1, $\\
    & with $i \in \{x,y,z\}$ & $\sigma_{\sigma_v} = 0.05$\\
    Flash shape & Correlated signal & See \hyperref[appendix:corr_matrix]{appendix}\\
    \hline
  \end{tabular}
\end{table}

The reconstruction of the position and movement, as well as the
characteristics of the bioluminescence lightcurve for four different random
seeds, are summarized in table~\ref{tab:recon_420_source_obs}.
For the reconstruction runs $2 - 4$ we observe similar results. 
The results of the remaining run $1$ deviate largely from the other ones,
especially the y and z source positions differ more than $5$ m. Even though,
the reconstructed observables of run $1$ are biologically plausible, the
residuals of run $1$ are larger compared to the runs $2 - 4$. Therefore, we
conclude that the runs $2-4$ yield better estimates.

The uncertainty estimates are significantly smaller than the deviations
between the reconstruction runs. As mentioned in section~\ref{sec:methods},
because of the mismatch between the assumed model and the real data generation,
the MGVI algorithm can not calculate reasonable uncertainty estimates. The
given uncertainties do not include systematic errors. Therefore, these
estimates are not given for the following data analysis, since they are not
representative.

\begin{figure}
  \includegraphics[width=0.95\columnwidth]{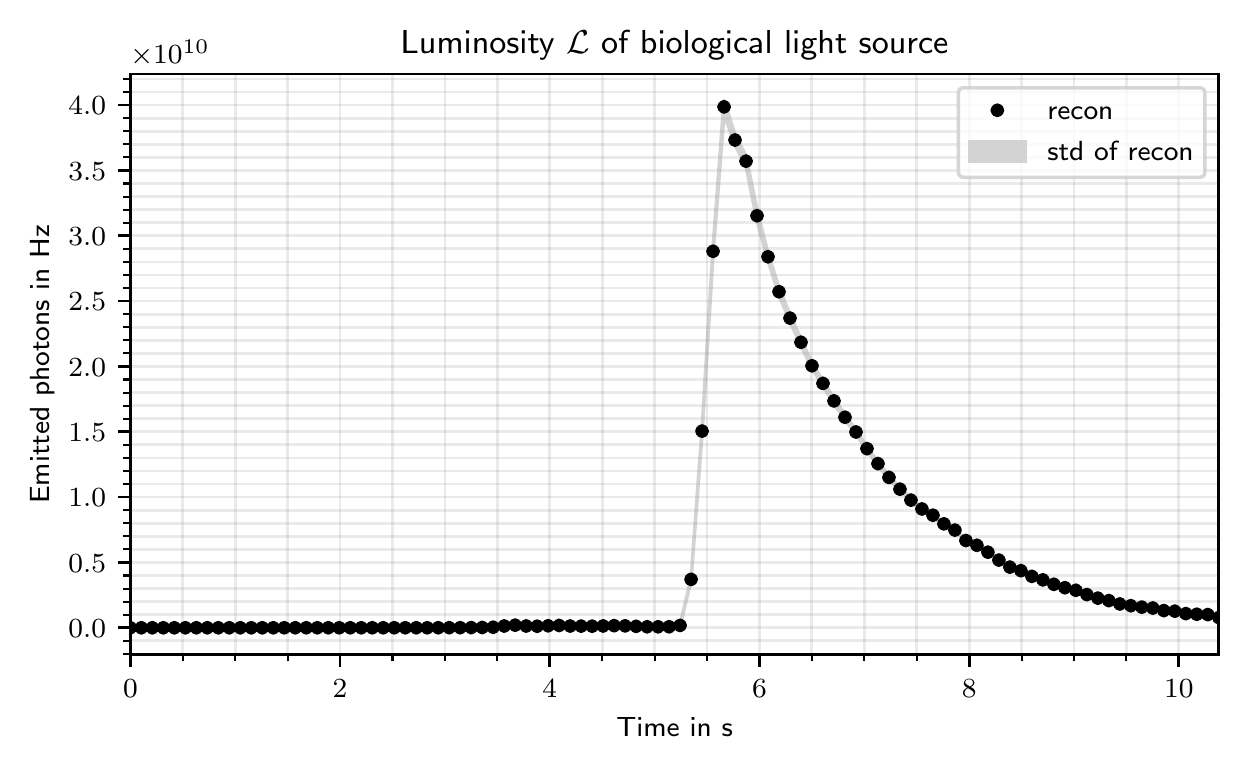}
  \caption{Reconstruction of the bioluminescence flash $1$ lightcurve detected by
  the ANTARES detector on the 11th of January 2010 at 04:12 (UTC).}
  \label{fig:420_lumin}
  \includegraphics[width=0.95\columnwidth]{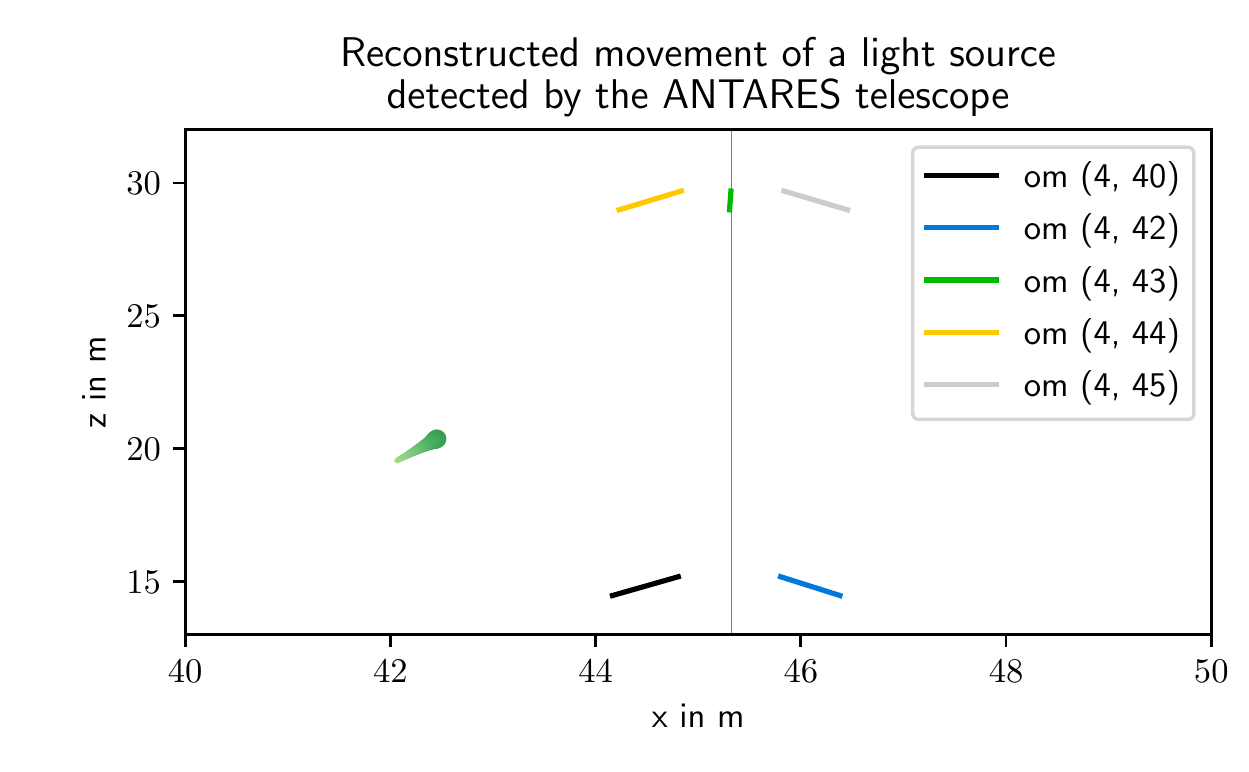}
  \caption{Movement reconstruction of a light source emitting flash $1$
  detected by the ANTARES detector on the 11th of January at 04:12 (UTC). 
  The light source is represented by overlapping green dots. The size of the dots 
  indicates the luminosity change over time without physical meaning. 
  The fading of the green color - from dark to bright - indicates the positive time flow.
  The optical module $(4, 41)$ on the bottom storey is not shown, since it malfunctioned.}
  \label{fig:420_xz}
\end{figure}

In figure~\ref{fig:420_lumin}, the reconstructed flash lightcurve of run $2$
is presented as it provides the smallest data residuals. We also illustrate the
estimated movement in figure~\ref{fig:420_xz}.

\subsection{ANTARES recordings of flash $2$}

A second flash was observed on the same day at 04:13:20 (UTC) recorded by optical
modules within the array $(4, 37 - 54)$. Flash $2$ peaked at around $85$~s
after the start of the recording shown in figure~\ref{fig:example_data} in section~\ref{sec:data}.
The optical modules $(4, 41)$ and
$(4, 51)$ did not detect the second flash as they already did not record counts
during the first flash. Furthermore, optical module $(4, 43)$ did not record
any photon for the period of highest luminosity, since the readout
electronics were saturated. All model parameters are the same as for flash
$1$ given in table~\ref{tab:recon_420_param}. The ANTARES detector is still
in a similar configuration, because the second flash occurred around one minute
later. Furthermore, due to similar photon count data of flash $1$ and flash
$2$, the same a priori assumption of the position can be taken.

The results of the different runs are summarized in table~\ref{tab:recon_860_source_obs}. 
Two different types of reconstructions can be
observed; run $2$ and $4$ reconstructed a rapid movement, whereas run $1$ and
$3$ estimate smaller changes in the position. Since the rapid movement for
run $2$ and $4$ occurs before the actual flash was emitted, no reasonable
estimate can be made for this period. When comparing the reconstructed
positions after the luminosity peak, we observe similar movements.

\begin{figure}
  \includegraphics[width=0.95\columnwidth]{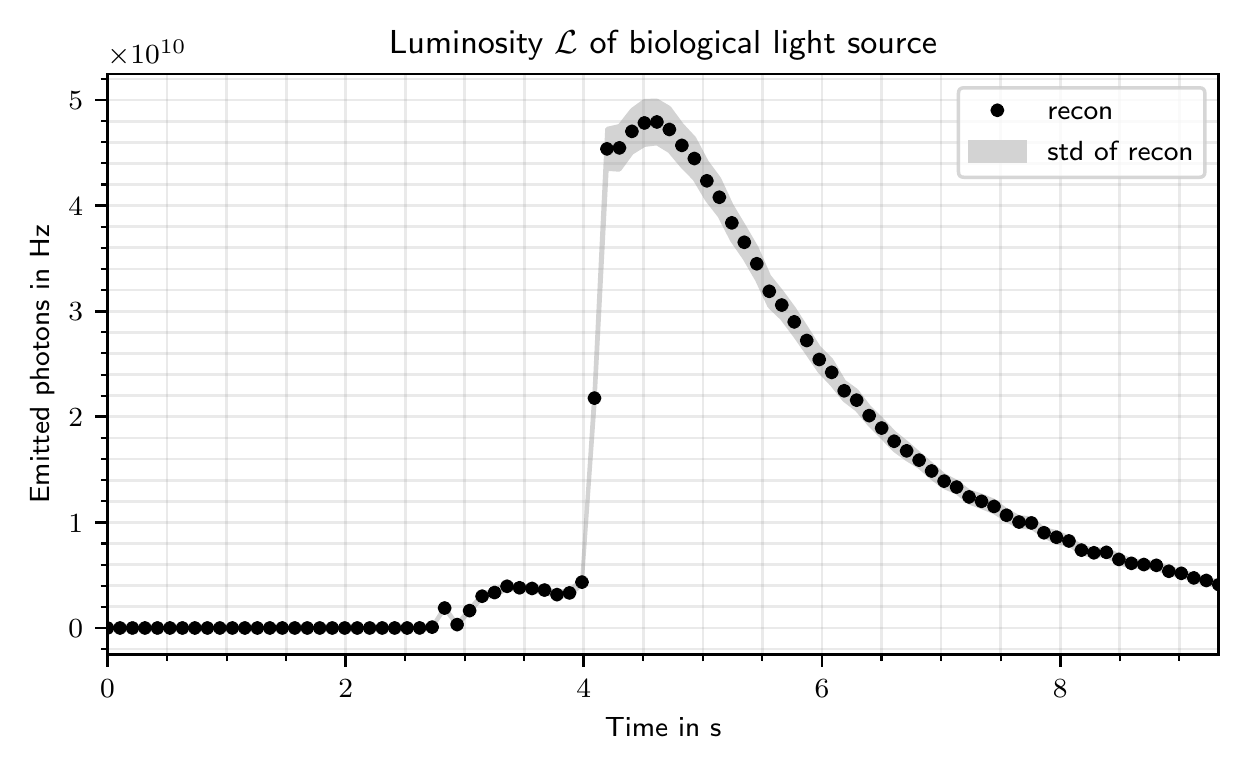}
  \caption{Reconstruction of the emitted bioluminescence flash $2$ lightcurve
  detected by the ANTARES detector on the 11th of January 2010 at
  04:13 (UTC).}
  \label{fig:860_lumin}
  \includegraphics[width=0.95\columnwidth]{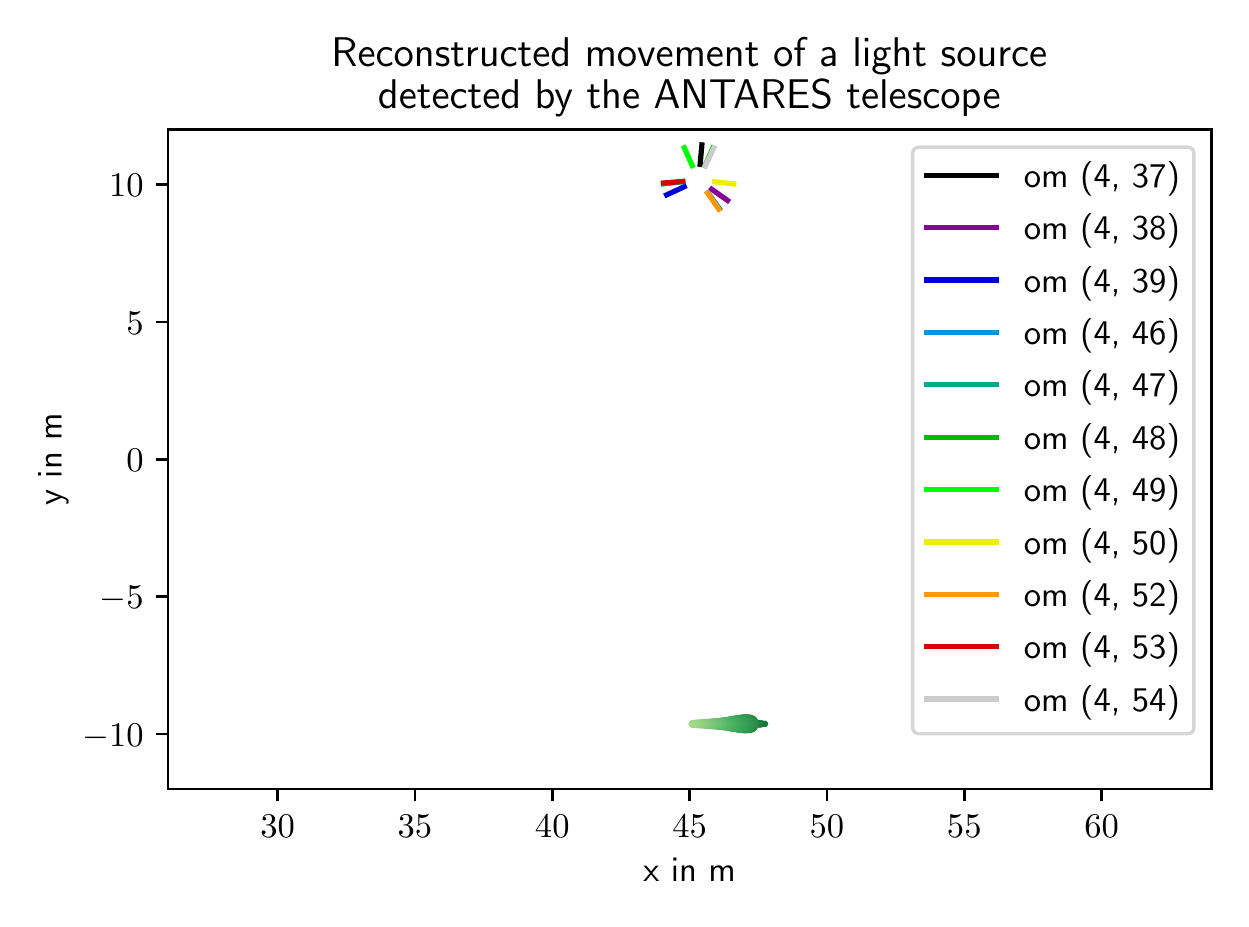}
  \caption{Movement reconstruction of a light source emitting flash $2$
  detected by the ANTARES detector on the 11th January at 04:13. 
  The light source is represented by overlapping green dots. The size of the dots 
  indicates the luminosity change over time without physical meaning. 
  The fading of the green color - from dark to bright - indicates the positive time flow.}
  \label{fig:860_xy}
\end{figure}

In figure~\ref{fig:860_lumin} we present the flash lightcurve of run $1$,
since the smallest data residuals could be calculated for this run. The
reconstructed movement of run $1$ is illustrated in figure~\ref{fig:860_xy}.

The second flash occurred only $44$~seconds after and around $13$~m away from
the first flash. A causal connection between both light emission is likely.
Due to the temporal and spatial closeness of the two events, it can not be
excluded that the light was emitted by the same organism. But the deviation
between the lightcurve structures of the two flashes also allow the
assumption of a recorded communication of two organisms.

\subsection{ANTARES recordings of flash $3$}

The last bioluminescence flash that is analyzed within this work occurred on
the 19th of January 2010 at 22:28 (UTC). In comparison to the previous models we
adjust the position model parameter since the flash was detected by different
optical modules. The position model parameters are given in table~\ref{tab:recon_110_param}. 
The remaining model parameters are the same as in
table~\ref{tab:recon_420_param}.

\begin{table}
  \caption{Model parameters for the initial position of a dynamic source. The
  model is used to reconstruct a biological source detected by the ANTARES
  telescope on the 19th of January 2010 at 22:28 (UTC) that emitted flash $3$. 
  The other model parameters are taken from table~\ref{tab:recon_420_param}.}
  \label{tab:recon_110_param} 
  \begin{tabular}{l l l} 
    \hline
    Observable & Model & Model parameter \\
    \hline
    $x$ position & Uniform & $x =(-25\text{ m}, 35\text{ m})$\\
    $y$ position & Uniform & $y=(65\text{ m}, 125\text{ m})$\\
    $z$ position & Uniform & $z=(-40\text{ m}, 10\text{ m})$\\ 
    \hline
  \end{tabular}
\end{table}

\begin{figure}[!h]
  \includegraphics[width=0.95\columnwidth]{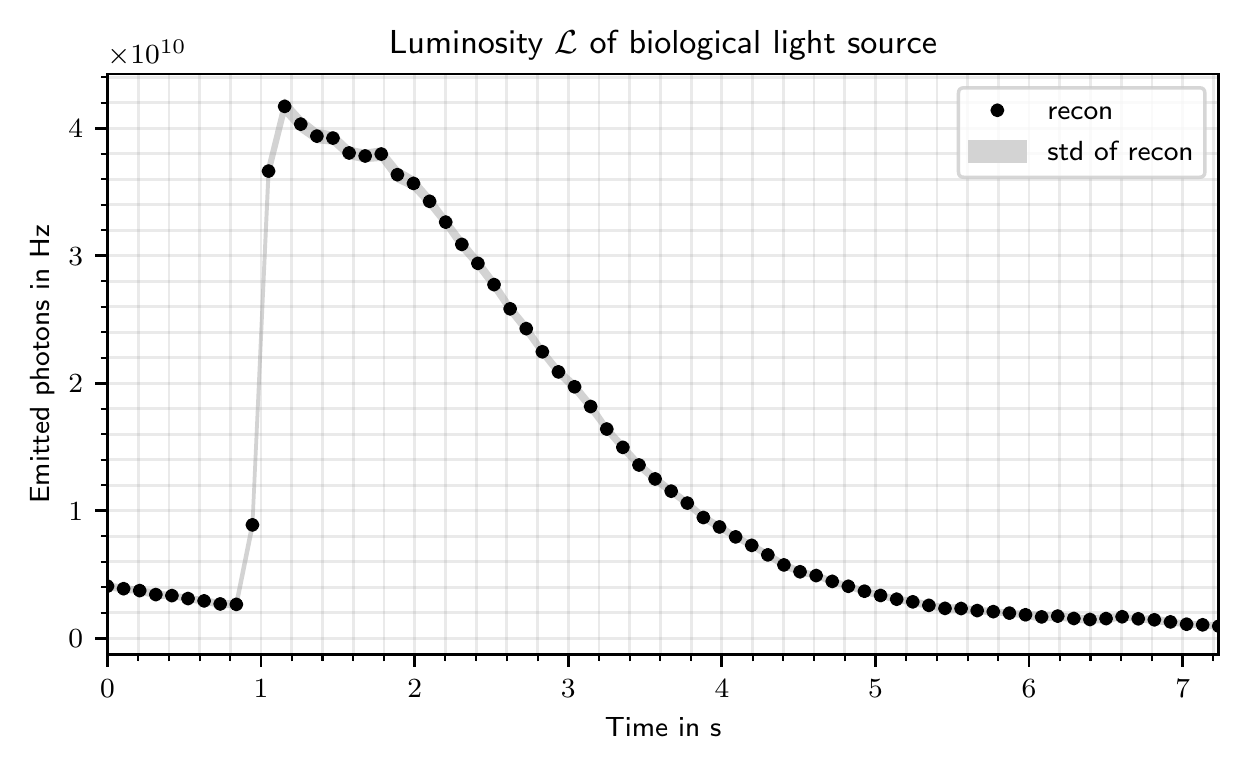}
  \caption{Reconstruction of the emitted bioluminescence flash $3$ lightcurve
  detected by the ANTARES detector on the 19th of January 2010 at 22:28 (UTC).}
  \label{fig:110_lumin}
  \includegraphics[width=0.95\columnwidth]{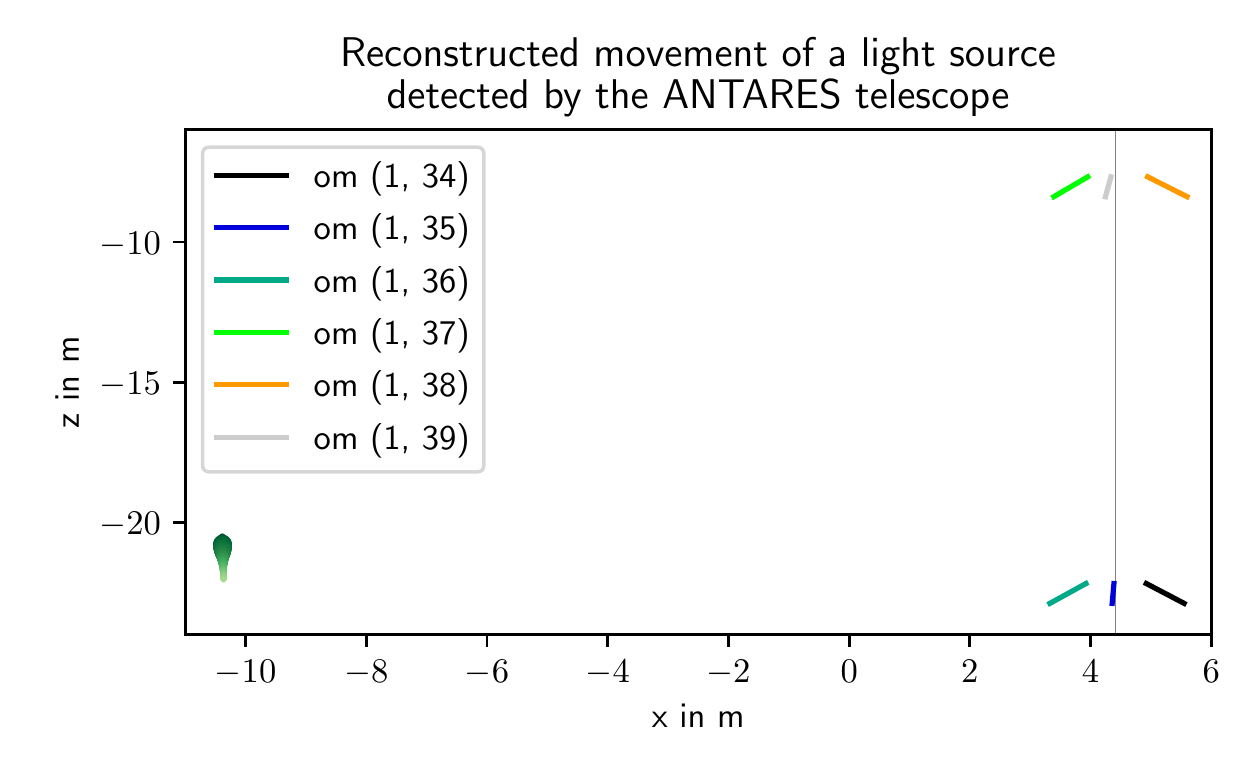}
  \caption{Movement reconstruction of a light source emitting flash $3$
  detected by the ANTARES detector on the 19th of January at 22:28 (UTC). 
  The light source is colored green. The size of the dots 
  indicates the luminosity change over time without physical meaning. 
  The fading of the green color - from dark to bright - indicates the positive time flow.}
  \label{fig:110_xz}
\end{figure}

Table~\ref{tab:recon_110_source_obs} summarizes the reconstruction runs and
clearly shows that run $1$ deviates from the other reconstructions. By
calculating the estimated velocity of the source as $3.8
\frac{\text{m}}{\text{s}}$ the result can be classified as biologically
implausible \cite{bio:deepseafish}. Besides that, the data residuals
are higher compared to the runs $2-4$, we hence can conclude that this run
resulted in a worse fit to the data. Therefore, as a reasonable reconstruction
result the flash lightcurve of run $2$ is presented in figure~\ref{fig:110_lumin}
and the reconstructed movement of this run illustrated in
figure~\ref{fig:110_xz}.

\begin{table*}[p]
  \caption{Reconstructed parameters assuming a dynamic light source model. The
  flash $1$ lightcurve used for the reconstruction was recorded on the 11th of
  January 2010 at 04:12 (UTC) by the ANTARES telescope. The positions are given at
  different times as in table~\ref{tab:recon_dynamic_dynamic_source_obs}.}
  \label{tab:recon_420_source_obs} 
  \begin{tabular}{l r r r r} 
    \hline
    Observable & Run 1 & Run 2 & Run 3 & Run 4 \\
    \hline
    $x(t_0)$ source position (m) & $45.35 \pm 0.00$ & $42.92 \pm 0.25$ & $42.21 \pm 0.23$ & $43.79 \pm 0.13$ \\
    $x(t_\text{h})$ source position (m) & $45.35 \pm 0.00$ & $42.45 \pm 0.12$ & $42.53 \pm 0.07$ & $42.6 \pm 0.09$ \\
    $x(t_\text{max})$ source position (m) & $45.35 \pm 0.00$ & $42.06 \pm 0.37$ & $41.71 \pm 0.14$ & $41.62 \pm 0.20$ \\
    $y(t_0)$ source position (m) & $10.88 \pm 0.06$ & $1.99 \pm 0.07$ & $2.19 \pm 0.15$ & $2.46 \pm 0.15$ \\
    $y(t_\text{h})$ source position (m) & $10.85 \pm 0.06$ & $2.00 \pm 0.05$ & $1.98 \pm 0.09$ & $2.05 \pm 0.06$ \\
    $y(t_\text{max})$ source position (m) & $10.82 \pm 0.07$ & $2.01 \pm 0.08$ & $1.81 \pm 0.18$ & $1.71 \pm 0.15$ \\
    $z(t_0)$ source position (m) & $15.00 \pm 0.00$ & $21.37 \pm 0.05$ & $21.25 \pm 0.24$ & $21.04 \pm 0.22$ \\
    $z(t_\text{h})$ source position (m) & $15.19 \pm 0.01$ & $20.37 \pm 0.05$ & $20.41 \pm 0.07$ & $20.42 \pm 0.07$ \\
    $z(t_\text{max})$ source position (m) & $15.14 \pm 0.01$ & $19.54 \pm 0.11$ & $20.07 \pm 0.12$ & $19.9 \pm 0.14$ \\
    \\
    Flash duration (s) & $5.14$ & $5.14$ & $5.14$ & $5.14$ \\
    $\mathcal{L}_\text{max}$ ($10^{10}$ Hz) & $4.2 \pm 0.02$ & $3.99 \pm 0.05$ & $3.97 \pm 0.04$ & $3.93 \pm 0.03$ \\
    Total emission ($10^{10}$ photons flash$^{-1}$) & $6.48$ & $5.66$ & $5.68$ & $5.62$ \\
    \hline
  \end{tabular}
\end{table*}

\begin{table*}[p]
  \caption{Reconstructed parameters assuming a dynamic light source. The flash $2$
  lightcurve used for the reconstruction was recorded on the 11th of January
  2010 at 04:13 (UTC) by the ANTARES telescope. The positions are given
  at different times as in table \ref{tab:recon_dynamic_dynamic_source_obs}.}
  \label{tab:recon_860_source_obs} 
  \begin{tabular}{l r r r r} 
    \hline
    Observable & Run 1 & Run 2 & Run 3 & Run 4 \\
    \hline
    $x(t_0)$ source position (m) & $48.90$ & $49.27$ & $48.81$ & $49.78$ \\
    $x(t_\text{h})$ source position (m) & $47.02$ & $47.21$ & $47.01$ & $47.42$ \\
    $x(t_\text{max})$ source position (m) & $45.10$ & $45.01$ & $45.17$ & $44.90$ \\
    $y(t_0)$ source position (m) & $-9.63$ & $-19.71$ & $-9.61$ & $-19.80$ \\
    $y(t_\text{h})$ source position (m) & $-9.63$ & $-10.62$ & $-9.61$ & $-13.15$ \\
    $y(t_\text{max})$ source position (m) & $-9.63$ & $-8.96$ & $-9.61$ & $-6.03$ \\
    $z(t_0)$ source position (m) & $19.46$ & $19.26$ & $19.56$ & $15.15$ \\
    $z(t_\text{h})$ source position (m) & $19.55$ & $19.59$ & $19.59$ & $19.04$ \\
    $z(t_\text{max})$ source position (m) & $19.65$ & $19.93$ & $19.62$ & $20.06$ \\
    \\
    Flash duration (s) & $6.61$ & $6.61$ & $6.61$ & $6.61$ \\
    $\mathcal{L}_\text{max}$ ($10^{10}$ Hz) & $4.79$ & $5.51$ & $4.81$ & $6.69$ \\
    Total emission ($10^{10}$ photons flash$^{-1}$) & $11.83$ & $13.09$ & $11.87$ & $14.27$ \\
    \hline
  \end{tabular}
\end{table*}

\begin{table*}[p]
  \caption{Reconstructed parameters assuming a dynamic light source model. The
  flash $3$ lightcurve used for the reconstruction was recorded on the 19th of
  January 2010 at 22:28 (UTC) by the ANTARES telescope. The positions are given at
  different times as in table \ref{tab:recon_dynamic_dynamic_source_obs}.}
  \label{tab:recon_110_source_obs} 
  \begin{tabular}{l r r r r} 
    \hline
    Observable & Run 1 & Run 2 & Run 3 & Run 4 \\
    \hline
    $x(t_0)$ source position (m) & $-24.58$ & $-10.38$ & $-11.09$ & $-10.39$ \\
    $x(t_\text{h})$ source position (m) & $-20.59$ & $-10.38$ & $-11.11$ & $-10.38$ \\
    $x(t_\text{max})$ source position (m) & $0.48$ & $-10.39$ & $-11.21$ & $-10.37$ \\
    $y(t_0)$ source position (m) & $110.44$ & $104.40$ & $104.81$ & $104.39$ \\
    $y(t_\text{h})$ source position (m) & $109.04$ & $-104.42$ & $104.85$ & $104.43$ \\
    $y(t_\text{max})$ source position (m) & $101.64$ & $104.51$ & $105.06$ & $104.63$ \\
    $z(t_0)$ source position (m) & $-23.77$ & $-20.60$ & $-20.31$ & $-20.54$ \\
    $z(t_\text{h})$ source position (m) & $-23.06$ & $-20.84$ & $-20.54$ & $-20.78$ \\
    $z(t_\text{max})$ source position (m) & $-19.29$ & $-22.10$ & $-21.77$ & $-22.03$ \\
    \\
    Flash duration (s) & $5.45$ & $7.34$ & $7.34$ & $7.34$ \\
    $\mathcal{L}_\text{max}$ ($10^{10}$ Hz) & $12.26$ & $4.17$ & $5.32$ & $4.18$ \\
    Total emission ($10^{10}$ photons flash$^{-1}$) & $18.55$ & $9.41$ & $12.31$ & $9.49$ \\
    \hline
  \end{tabular}
\end{table*}

\section{Discussion}

The application of the method on synthetic and real data reveals both the
strengths and the limitations of our approach. The reconstruction of the flash
lightcurve is limited by the remaining degeneracy between the efficiencies
and attentuation length with the emitted photon numbers. Therefore,
increasing the accuracy of the measurements of these quantities
will allow more precise reconstructions of the total
number of emitted photons.

Furthermore, the accuracy of the position reconstructions and the spatial
resolution can be increased by reducing the angular acceptance and increasing
the number of OMs. An OM with sharp angular acceptance is able to resolve
small position changes more precisely, since changes of the angle between OMs
and light source lead to significant changes in the arriving photon numbers.
However, a small angular acceptance leads also to a smaller monitored volume,
which can be compensated by using more OMs.

Although only a small number of photon data recorded by the ANTARES telescope
has been analyzed in this study, the results already provide some insights on
the biological sources. Our localizations of bioluminescence events showed
that the light emitting organisms were sufficiently far from optical modules, so
that they can be assumed to be undisturbed by them. Further measurements
of the environment, e.g. sea current measurements, need to be taken into account
to verify the occurence of non-stimulated bioluminescence events. According
to previous studies non-stimulated light emissions are rare 
\cite{bio:freefall, bio:monterey, bio:relation, bio:blooms} and hardly distinguishable
from stimulated ones using moving detectors \cite{bio:freefall}. Therefore, our method
opens the possibility to study spontaneous and in-situ bioluminescence. 
Further analysis of data recorded by deep sea neutrino telescopes could give insights 
about the frequency and distribution of such spontaneous events and the occurrence of 
consecutive light emissions as already observed in this work.

\section{Conclusion}

This work shows the potential of bioluminescence studies with a neutrino
telescope in the deep sea and highlights the biological activity information
that can be extracted. But it also clearly points out the limitations of the
bioluminescence studies in regards to the spatial resolution due to the
architecture of the detector.

The proposed method is generic and can be applied on data sets of different
underwater neutrino telescopes. The development and work on the new neutrino
telescope in the Mediterranean Sea, KM3NeT \cite{km3net:2016}, 
offers a detector architecture which is even
more suitable for the study of luminescent organisms. Each optical module of
KM3NeT will be equipped with $31$ photomultiplier tubes having a narrower
angular acceptance compared to the ANTARES setup \cite{km3net:2016}. This
increases the spatial resolution of the positioning.

For future systematic surveys of bioluminescence the method needs to be
automatized and optimized. An optimized framework can be used to build a
catalogue of various types of bioluminescence lightcurves including the
position of the source. In this work, we also showed that the tracking of
light sources is possible. Therefore, this method can also be used to analyze
the movement behavior of deep sea organisms, which still little is known
about.

\section*{Appendix}

\subsection*{Distribution transformation}
\label{appendix:distr_trans}

Since the MGVI algorithm relies on standardized variables and therefore, the
generative model is built on these variables, it is crucial to transform a
Gaussian distribution with zero mean and unit covariance to specific
distributions which are more reasonable for the mathematical model. Within this
work the common transformations are between standardized Gaussian
distributions and Gaussian distributions with given mean and covariance and
between standardized Gaussian distributions and uniform distributions.
The standardized variables are introduced as
\begin{align}
  \vec{\xi} \hookleftarrow \mathcal{G}(\vec{\xi},\mathbb{I}).
\end{align}
The transformation $s(\xi)$ from $\vec{\xi}$ to $\vec{s}$ sampled from a
Gaussian distribution with given mean $\vec{\mu}$ and correlation matrix $C$ is
given as
\begin{align}
  \vec{s} = s(\xi) = \vec{\mu} +  \sqrt{C} \cdot \vec{\xi}.
  \label{eq:gaussian_sampling}
\end{align}
Reducing the dimension to scalar values and also introducing a Gaussian
distributed standard deviation $\sigma \hookleftarrow \mathcal{G}(\sigma -
\overline{\sigma}, \sigma_\sigma^2)$ gives the new transformation with Gaussian
sampled standard deviation
\begin{align}
  s(\xi_\sigma, \xi) &= \mu + \sigma \cdot \xi \\
  &= \mu + \underbrace{(
  \overline{\sigma} + \sigma_{\sigma} \cdot \xi_\sigma)}_{\sigma \hookleftarrow
  \mathcal{G}(\sigma - \overline{\sigma}, \sigma_{\sigma}^2)} \cdot \xi
\end{align}\\
A transformation from standardized variables to uniformly sampled variables $s
\hookleftarrow \mathcal{U}([t_0, t_1])$ can be done by using the cumulative
distribution function (CDF) of a Gaussian distribution. Variables uniformly
sampled from the range $[0,1]$ can be generated by the transformation
\begin{align}
  s([0,1], \xi) = \text{CDF}(\xi).
\end{align}
By shifting and expanding, the standard uniform distribution can be transformed
into any uniform distribution with range $[t_0, t_1]$
\begin{align}
  s([t_0,t_1], \xi) = (t_1 - t_0) \cdot \text{CDF}(\xi) + t_0.
\end{align}

\subsection*{Correlation matrix for luminosity bursts}
\label{appendix:corr_matrix}

According to the Wiener-Khintchin Theorem, a statistical homogeneous and
isotropic signal $s$ drawn from a Gaussian prior $\mathcal{G}(s-\mu_s, S)$
with stationary auto-correlation $S^{t_it_j}= S^{ij} = C_s(t_i - t_j)$ becomes
diagonal in Fourier space,

\begin{align}
  S^{ij} = (2 \pi)^u \delta(w-w') C_s(w),
\end{align}
with $\Delta = t_i - t_j$ and $C_s(w) = \int d \Delta e^{iw \Delta} C_s(\Delta)$
as the Fourier transformed auto-correlation function. $C_s(w)$ is identical
to the power spectrum per time length $T$, \\$P_s(w) = \lim_{T \rightarrow \infty}
\frac{1}{T} \Big \langle | \int_T dt s^t e^{iwt}|^2 \Big \rangle = C_s(w)$
\cite{ift:niftysoftware}.

Therefore, the covariance can be described efficiently by defining a power
spectrum per time length $T$. In the following the power spectrum used to sample
the luminosity bursts is discussed. Recall, that the signal $s$ sampled with a
given power spectrum is exponentiated to produce the bursts. 

Instead of an inferable power spectrum, as used for the velocity vector, a fixed
power spectrum has shown to be sufficient to reproduce the burst shapes and
this assumption also reduces the computation time.

Power spectra of the form 
\begin{align}
  P_s(w) = [A_s(\omega)]^2 = \frac{b}{\omega^p}
  \label{eq:ps}
\end{align}
%P_s(w) = [A_s(w)]^2 = \Big [ \frac{1}{T \cdot \sqrt{\sum_{k>0} k^2}}  \frac{b}{w^p} \Big ]^2
are used within this work with $A_s(\omega)$ defined as amplitude. The amplitude
is divided into two parts, the zero mode $A_0$ and the remaining part
$A_s(\omega)$ with $\omega > 0$. Intuitive parameters $a_0, b, p$ are
introduced. Following relations ensure definitions of power spectra to be
independent of the size of the total time length $T$. The zero mode $A_0$ is
defined by a relative zero mode $a_0$ and given as
\begin{align}
  A_0 = \frac{a_0}{T}.
\end{align}
The remaining part of the amplitude is defined by the power law $p$ and an
amplitude factor $b$. The relation is given by
\begin{align}
  A_s(\omega) = \frac{b}{T \sqrt{\sum_{\omega' \neq 0}r(\omega')^2}} \cdot
r(\omega)
\end{align}
with $r(\omega) = \frac{1}{\omega^p}$ and $\omega > 0$. The parameters $a_0, b$
and $p$ are set by the user.

The table \ref{tab:pslumin} provides a summary of variables
defining the luminosity burst shape. Their use is briefly explained and
the values used within this work are given.

\begin{table}[!h]
  \centering
  \caption{Summary of parameters used for the correlated log-signal of luminosity bursts.}
  \begin{tabular}{l l l}
    \hline 
    Parameter & Value & Explanation \\
    \hline
    Amplitude factor $b$ & $8.0$  & Changes the magnitude \\
    & & of the fluctuations \\
    Power law $p$ & $1.5$ & Changes the smoothness \\
    & & of the burst \\
    Relative zero mode $a_0$ & $2.0$ & Coupled to the variance\\
    & & of the mean $\mu_s$ \\
    Mean $\mu_s$ & $4.0$ & Mean of the signal \\ 
    \hline
  \end{tabular}
  \label{tab:pslumin}
\end{table}

%%%%%%%%%%%%%%%%%%%%%%%%%%%%%%%%%%%%%%%%%%%%%%
%%                                          %%
%% Backmatter begins here                   %%
%%                                          %%
%%%%%%%%%%%%%%%%%%%%%%%%%%%%%%%%%%%%%%%%%%%%%%

\begin{backmatter}

\section*{Acknowledgements}%% if any
SH acknowledges funding from the European Research Council (ERC) under the
European Union's Horizon 2020 research and innovation programme (grant
agreement No 772663)

The authors acknowledge the financial support of the funding agencies:
% France:
Centre National de la Recherche Scientifique (CNRS), Commissariat \`a
l'\'ener\-gie atomique et aux \'energies alternatives (CEA),
Commission Europ\'eenne (FEDER fund and Marie Curie Program),
Institut Universitaire de France (IUF), LabEx UnivEarthS (ANR-10-LABX-0023 and ANR-18-IDEX-0001),
R\'egion \^Ile-de-France (DIM-ACAV), R\'egion
Alsace (contrat CPER), R\'egion Provence-Alpes-C\^ote d'Azur,
D\'e\-par\-tement du Var and Ville de La
Seyne-sur-Mer, France;
% Germany: 
Bundesministerium f\"ur Bildung und Forschung
(BMBF), Germany; 
% Italy
Istituto Nazionale di Fisica Nucleare (INFN), Italy;
% Netherlands
Nederlandse organisatie voor Wetenschappelijk Onderzoek (NWO), the Netherlands;
% Russia
Council of the President of the Russian Federation for young
scientists and leading scientific schools supporting grants, Russia;
% Romania
Executive Unit for Financing Higher Education, Research, Development and Innovation (UEFISCDI), Romania;
% Spain
Ministerio de Ciencia, Innovaci\'{o}n, Investigaci\'{o}n y Universidades (MCIU): Programa Estatal de Generaci\'{o}n de Conocimiento (refs. PGC2018-096663-B-C41, -A-C42, -B-C43, -B-C44) (MCIU/FEDER), Generalitat Valenciana: Prometeo (PROMETEO/2020/019), Grisol\'{i}a (ref. GRISOLIA/2018/119) and GenT (refs. CIDEGENT/2018/034, /2019/043, /2020/049) programs, Junta de Andaluc\'{i}a (ref. ref. A-FQM-053-UGR18), La Caixa Foundation (ref. LCF/BQ/IN17/11620019), EU: MSC program (ref. 101025085), Spain;
% Marocco
Ministry of Higher Education, Scientific Research and Professional Training, Morocco.
% A.O.B.:
We also acknowledge the technical support of Ifremer, AIM and Foselev Marine
for the sea operation and the CC-IN2P3 for the computing facilities.

%\section*{Funding}%% if any
%Text for this section\ldots
%
%\section*{Abbreviations}%% if any
%Text for this section\ldots
%
%\section*{Availability of data and materials}%% if any
%Text for this section\ldots
%
%\section*{Ethics approval and consent to participate}%% if any
%Text for this section\ldots
%
%\section*{Competing interests}
%The authors declare that they have no competing interests.
%
%\section*{Consent for publication}%% if any
%Text for this section\ldots
%
%\section*{Authors' contributions}
%Text for this section \ldots
%
%\section*{Authors' information}%% if any

%%  After submission of the .TEX file,                     %%
%%  you will be prompted to submit your .BBL file.         %%

% if your bibliography is in bibtex format, use those commands:
\bibliographystyle{bmc-mathphys} % Style BST file (bmc-mathphys, vancouver, spbasic).
\bibliography{article_antares}      % Bibliography file (usually '*.bib' )
% for author-year bibliography (bmc-mathphys or spbasic)
% a) write to bib file (bmc-mathphys only)
% @settings{label, options="nameyear"}

% or include bibliography directly:
% \begin{thebibliography}
% \bibitem{b1}
% \end{thebibliography}

%%%%%%%%%%%%%%%%%%%%%%%%%%%%%%%%%%%
%%                               %%
%% Figures                       %%
%%                               %%
%% NB: this is for captions and  %%
%% Titles. All graphics must be  %%
%% submitted separately and NOT  %%
%% included in the Tex document  %%
%%                               %%
%%%%%%%%%%%%%%%%%%%%%%%%%%%%%%%%%%%

\end{backmatter}
\end{document}